\documentclass[10pt]{article}
\usepackage{sao2}
\usepackage{psfig,epsf}

\issue{2010, 65, 390--399}

\def\apj{Astrophys. J}

\def\apjs{Astrophys. J. Supp.}

\begin{document}
\markboth{O.V.~Verkhodanov,M.L.~Khabibullina}
 {DOMINANT MULTIPOLES IN WMAP5 DATA CORRELATIONS}
\title{Dominant Multipoles in WMAP5 Mosaic Data Correlation Maps}
\author{O.V.~Verkhodanov\inst{a},
M.L.~Khabibullina\inst{a}
}
\institute{
$^a$\saoname}

\date{November 30, 2010}{March 31, 2010}
\maketitle

\begin{abstract}
The method of correlation mapping on the full sphere is used to
study the properties of the ILC map, as well as the dust and
synchrotron background components. An anomalous correlation of the
components with the ILC map in the main plane and in the poles of
the ecliptic and equatorial coordinate systems was discovered.
Apart from the bias, a dominant quadrupole contribution in the
power spectrum of the mosaic correlation maps was found in the
pixel correlation histogram. Various causes of the anomalous
signal are discussed.
\end{abstract}

\maketitle

\section{INTRODUCTION}

Non-Gaussian features for the low harmonics, the quadrupole and
octupole, were found almost immediately after the first WMAP
mission data \mbox{release
\cite{wmapresults:Verkh_ecl_en,wmapfg:Verkh_ecl_en,wmappara:Verkh_ecl_en}.}
These features are expressed in the planarity and alignment of the
two harmonics, which was noted by Tegmark et al.
\cite{toh:Verkh_ecl_en}. The axis along which the extrema of
these multipoles are lined up in the WMAP mission map was later
named the Axis of \mbox{Evil \cite{axisevil:Verkh_ecl_en}.} Note
that this axis is not the only selected axis in the mission data,
the naturally selected axis lies in the Galactic plane. In the
further releases of the WMAP mission data
\cite{wmap3ytem:Verkh_ecl_en,wmap3ycos:Verkh_ecl_en,wmap5ytem:Verkh_ecl_en,wmap5ycos:Verkh_ecl_en}
the existence of the Axis of Evil was also confirmed.

Different estimations of the significance of existence of this
axis, and several hypotheses on its origin were made. Various
studies, e.g. \cite{copi2006:Verkh_ecl_en,gruppuso2009:Verkh_ecl_en}~ investigated the
contribution of background components and their influence on the
alignment of multipoles ($\ell=2$ and $\ell=3$), and indicated a
small probability of the background effect on the orientation of
the low multipoles. In \cite{copi2006:Verkh_ecl_en}, where the multipole
vectors were used for the estimates of this effect, it was also
noted that the positions of the quadrupole and octupole axes
correspond to the geometry and direction of motion of the Solar
System and are perpendicular to the ecliptic plane and the plane,
given by the direction to the dipole. Randomness of such an effect
is estimated by the authors as unlikely at the	    significance
level exceeding 98\% and exclude the effect of residual
contribution of background components. Continuing the research
done, Copi et al. \cite{copi2009:Verkh_ecl_en} conclude that the
characteristics of low multipoles are abnormally different from
random, which may be due to the statistical anisotropy of the
universe at large scales, or to the problems of the ILC (Internal
Linear Combination) signal deconvolution method. Park et al.
\cite{park2007:Verkh_ecl_en} note that the planarity of the quadrupole and
octupole is not statistically significant. They also stress that
the residual photon radiation in the ILC map does not affect
significantly the level of the effect.

Cosmological models were developed to explain the prominence of
the axis in the orientation of multipoles. The alignment of the
quadrupole and octupole could be explained within the framework of
these models. Various models include the anisotropic expansion of
the Universe, rotation and magnetic \mbox{field
\cite{jaffe2006:Verkh_ecl_en,dem_dor:Verkh_ecl_en,koivisto:Verkh_ecl_en}.}

Other studies investigated the contribution of ecliptic dust in
the microwave background. Diego et al. \cite{diego:Verkh_ecl_en} have
constructed a combination of initial data of the WMAP channels:
$V+W-2Q$,  which was smoothed over a 7-degree diagram. The CMB is
not included in this combination, however, it can be used as a
tool to investigate the behavior of the large-scale residual
signal. The authors have shown the presence of radiation in this
residual signal. Its distribution is correlated with the position
of the ecliptic plane. The paper investigates  the contribution of
zodiacal dust and shows that the radiation level of this
contribution is not sufficient to explain the level of residual
signal in the map combinations. In addition, a contribution from
unresolved sources was estimated, which could also become an
explanation of the observed effect. The authors do not consider
this contribution sufficient as well.

Dikarev et al. \cite{dikarev:Verkh_ecl_en} used the WMAP data and the infrared
survey data to compute the parameters of dust particles, which
could provide the corresponding contribution in the microwave
background. The authors believe that the silicate beads a few
millimeters in size in the zodiacal cloud and trans-Neptunian
region with optical depth on the order of 10$^{-7}$ could explain
additional radiation at the level of 10\,$\mu$K in the microwave
background in the ecliptic.

We investigated the problem of correlation properties of the WMAP
maps, given the fact \cite{cormap:Verkh_ecl_en} that increased correlations
were previously reported  in various ILC signal distribution
regions. In addition, it was proved \cite{ndv03:Verkh_ecl_en,ndv04:Verkh_ecl_en,nv_quad:Verkh_ecl_en}
that a great contribution to the non-Gaussianity of data on
various multipoles is given by the background components. Coming
back to \cite{cormap:Verkh_ecl_en}, let us note that the distribution of the
correlation coefficients in mosaic areas of the map had a great
shift for the correlations with the dust component of the
background. This may be due to a more complex model of the dust
distribution than the one currently used  for the separation of
components, and then the effect may occur in the correlation at
other galactic latitudes.

In this paper, we continue to study the detected correlation
\cite{cormap:Verkh_ecl_en}  at different angular  scales. To study the effect,
we calculate the angular power spectrum of the mosaic correlation
map and select its main harmonic components. We apply this
analysis to the ILC map correlations both with the dust component,
and with the synchrotron radiation.

\section{MOSAIC CORRELATION METHOD}

To analyze the map's properties on different angular scales, we
expand the signal distributed on the sphere into the spherical
harmonics (multipoles):

\begin{equation}
\Delta S(\theta,\phi)= \sum_{\ell=1}^{\infty}\sum_{m=-\ell}^{m=\ell}
       a_{\ell m} Y_{\ell m} (\theta, \phi)\,,
\label{eq1:Verkh_ecl_en}
\end{equation}

\noindent where $\Delta S(\theta,\phi)$ are the signal variations
on the sphere in polar coordinates, $\ell$ is the multipole
number, $m$ is the multipole mode number. The spherical harmonics
are determined as
\begin{eqnarray}
Y_{\ell m}(\theta,\phi) = \sqrt{{(2\ell+1)\over 4\pi}{(\ell-m)!
\over (\ell+m)!}}P_\ell^m(x) e^{i m\phi}, \label{eq2:Verkh_ecl_en}
\end{eqnarray}
where $x=\cos\theta$, and $P_\ell^m(x)$ are the associated
Legendre polynomials. For a continuous function $\Delta S(x,\phi)$
the expansion coefficients $a_{\ell m}$ are expressed as
\begin{equation}
a_{\ell m}=\int^1_{-1}dx\int^{2\pi}_0 \Delta S(x,\phi)
Y^{*}_{\ell m}(x,\phi) d\phi\,,
\label{eq3:Verkh_ecl_en}
\end{equation}
where $Y^{*}_{\ell m}$ denotes a complex conjugation  $Y_{\ell
m}$.

The correlation properties of two maps of the signal distribution
on the sphere can be described, on a given angular scale, by a
correlation coefficient for the corresponding multipole $\ell$ as:
$$
K(\ell) = \frac{1}{2}
\frac{\sum\limits_{m=-\ell}^l t_{\ell m}s^*_{\ell m} + t^*_{\ell m}s_{\ell m}}
       {(\sum\limits_{m=-\ell}^l |t_{\ell m}|^2
     \sum\limits_{m=-\ell}^\ell |s_{\ell m}|^2)^{1/2}}\,,
$$
where $t_{\ell m}$ and $s_{\ell m}$ are the variations of two
signals in a harmonic representation. The coefficient $K(\ell)$
can be used to assess the correlation between the harmonics on the
sphere, i.e., to compare the properties of maps on a given angular
scale. However, in the case of a search for correlated areas,
which do not repeat in other regions of the sphere, this approach
smears such single areas in the process of averaging over the
sphere within a certain harmonic. In this case it becomes
practically impossible to identify the correlated areas.

Verkhodanov et al.~\cite{cormap:Verkh_ecl_en} proposed an approach, which was
implemented in the second release of the GLESP package
\cite{glesp2:Verkh_ecl_en}. The proposed procedure makes it possible to find
correlations between two maps in the areas of a certain angular
size. In this method, each pixel with number $p$ subtending the
solid angle  $\Xi_p$ is assigned the cross-correlation coefficient
$K(\Xi_p|\ell_{max})$ between the data of the two maps on the
corresponding area. Thus a correlation map is constructed for two
signals	 $T$ and $S$, where the value of each pixel $p$
($p=1,2,...,N_0$, \mbox{and $N_0$ is the} total number of pixels
on the sphere) with the subtending angle $\Xi_p$, and computed for
the sphere maps with the initial resolution determined by
$\ell_{max}$ is \mbox{equal to}

\small{\begin{eqnarray}
K(\Xi_p|\ell_{max}) =  \hspace{5cm} \nonumber \\
\frac{\sum\sum\limits_{p_{ij}\in\Xi_p}
	   (T(\theta_i,\phi_j) - \overline{T(\Xi_p))}
	   (S(\theta_i,\phi_j) - \overline{S(\Xi_p))}}
	{\sigma_{T_p}\sigma_{S_p}}.
\end{eqnarray}}
\normalsize

\noindent Here $T(\theta_i,\phi_j)$ is the value of the signal $T$
in the pixel with the coordinates $(\theta_i,\phi_j)$ for the
initial resolution of the pixelization of the sphere;
$S(\theta_i,\phi_j)$ is the value of the other signal in the same
area; $\overline{T(\Xi_p)}$ and \mbox{$\overline{S(\Xi_p)}$ } are
the mean values averaged over the area $\Xi_p$ and obtained from
the higher resolution map data determined by $\ell_{max}$,
$\sigma_{T_p}$, and $\sigma_{S_p}$ are the corresponding standard
deviations in the area considered.

\section{ILC AND DUST BACKGROUND COMPONENT MAP CORRELATION}

In \cite{cormap:Verkh_ecl_en} we found a significant bias in the
distribution of the correlation coefficients for the WMAP ILC and
dust maps. Let us view this distribution in detail. To do this, we
construct mosaic correlation maps with pixels having the sides of
160, 300 and 540 minutes of arc with the corresponding maximal
multipoles $\ell_{max}=33$, 17 and 9 in accordance with the
definition (4). The resulting maps are shown in Fig.\,1.

\begin{figure} 
\centerline{\vbox{
\psfig{figure=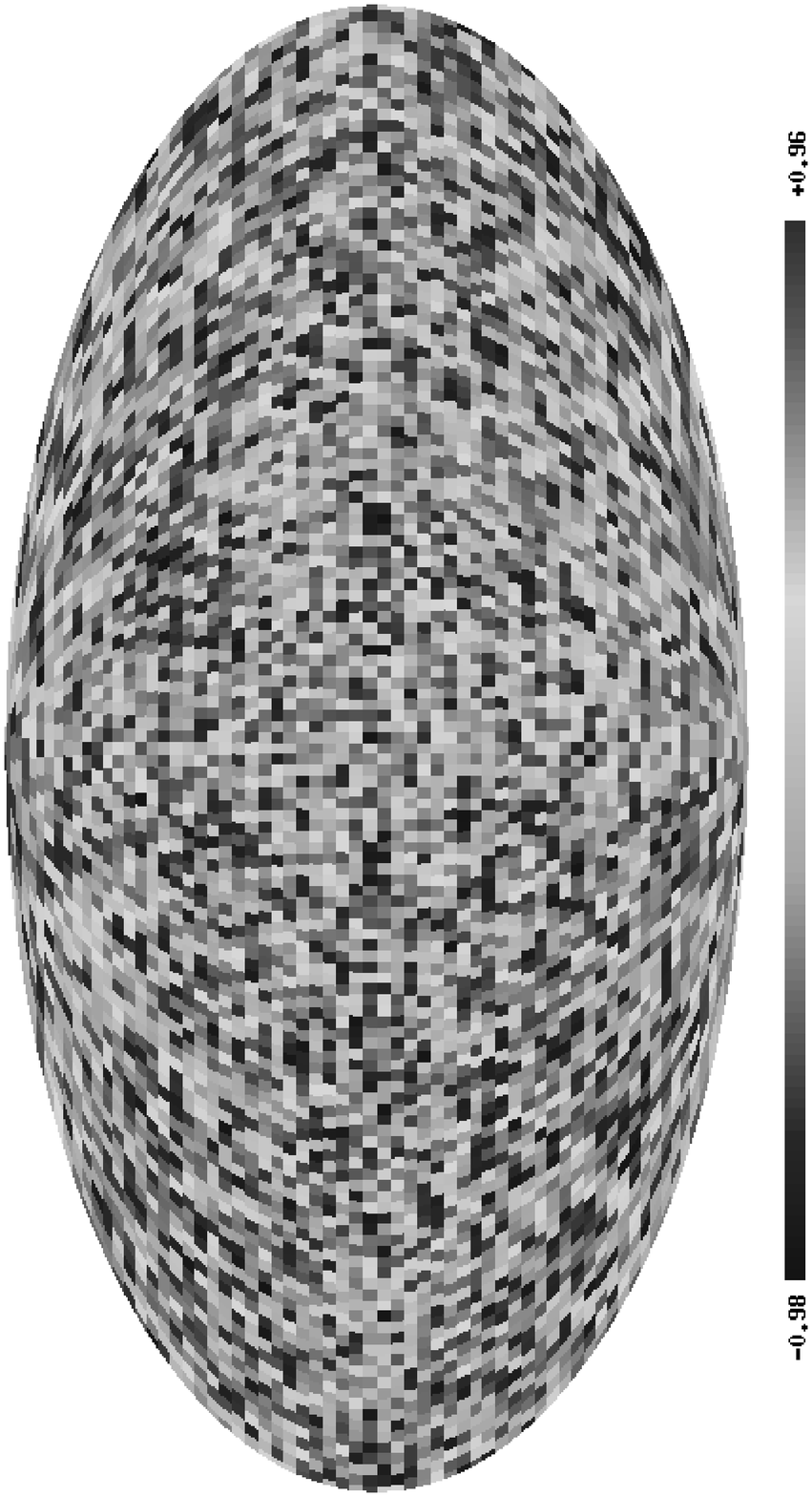,angle=-90,width=9cm}
\psfig{figure=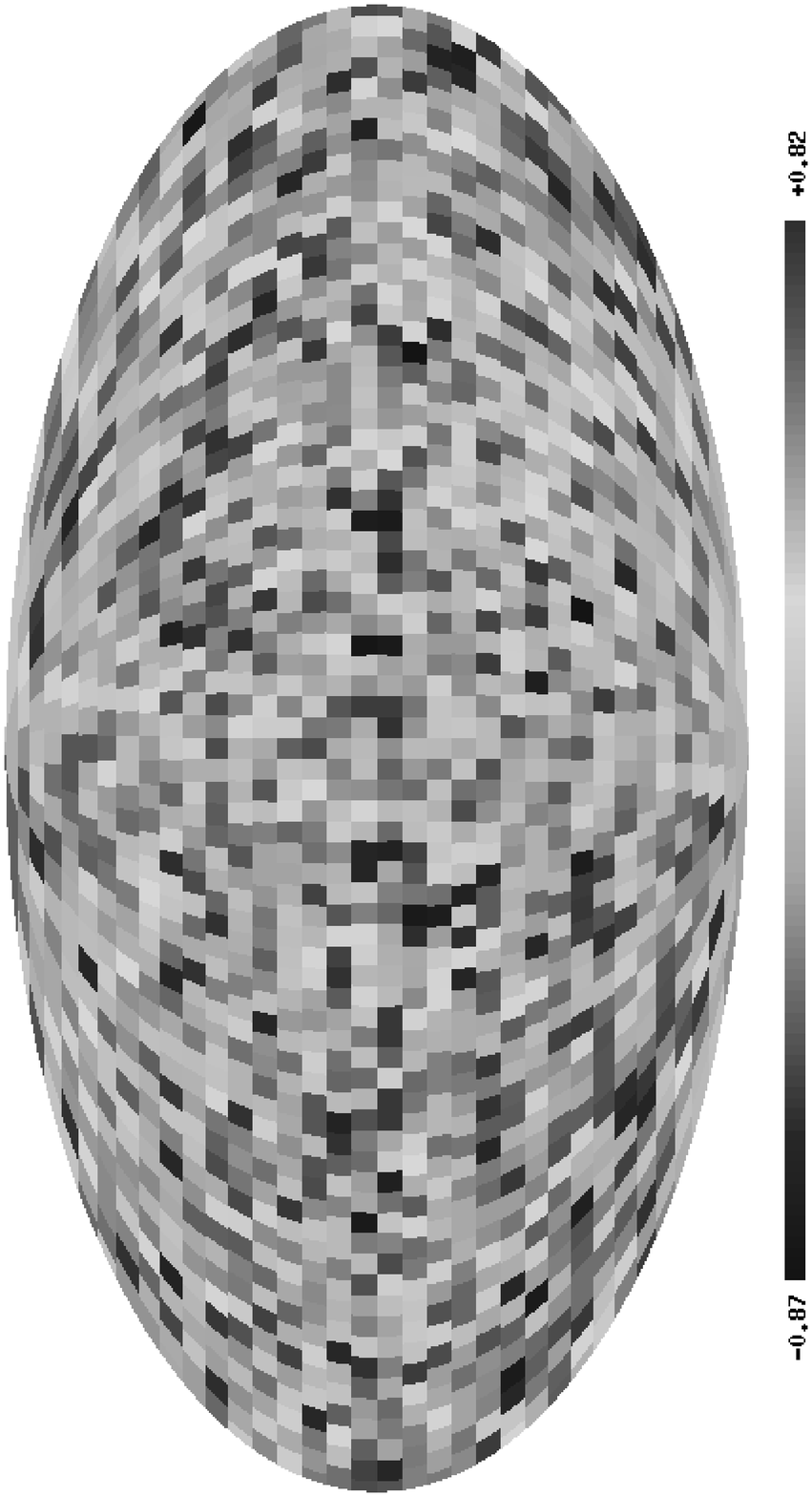,angle=-90,width=9cm}
\psfig{figure=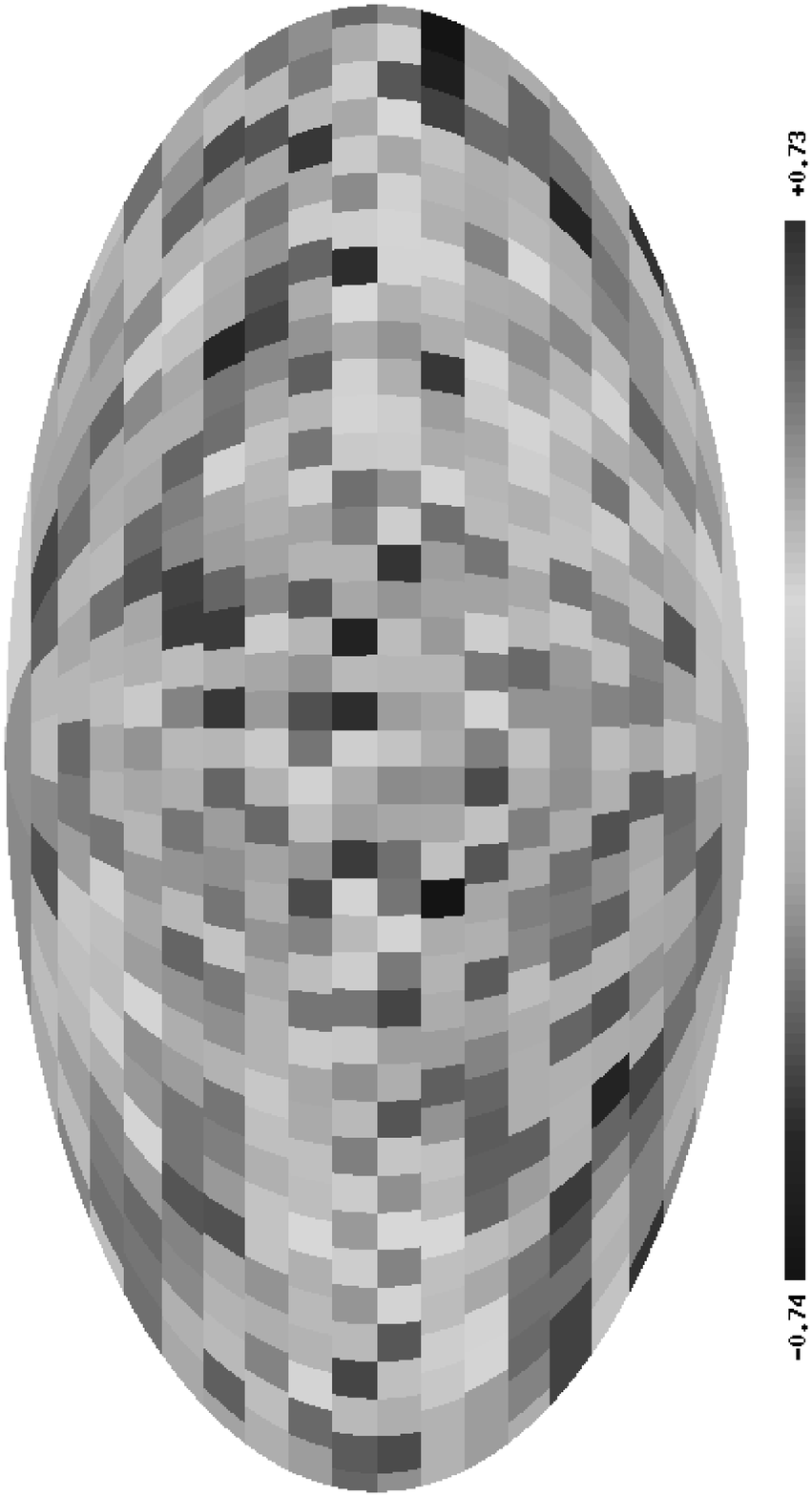,angle=-90,width=9cm}
}} \caption{Maps of correlation coefficients on the sphere,
constructed for mosaic correlations of the ILC and dust maps. Top
to bottom: maps for the pixelization with the pixel side of 160\arcmin,
300\arcmin and 540\arcmin.}
\end{figure}

\begin{figure} 
\centerline{\vbox{
\psfig{figure=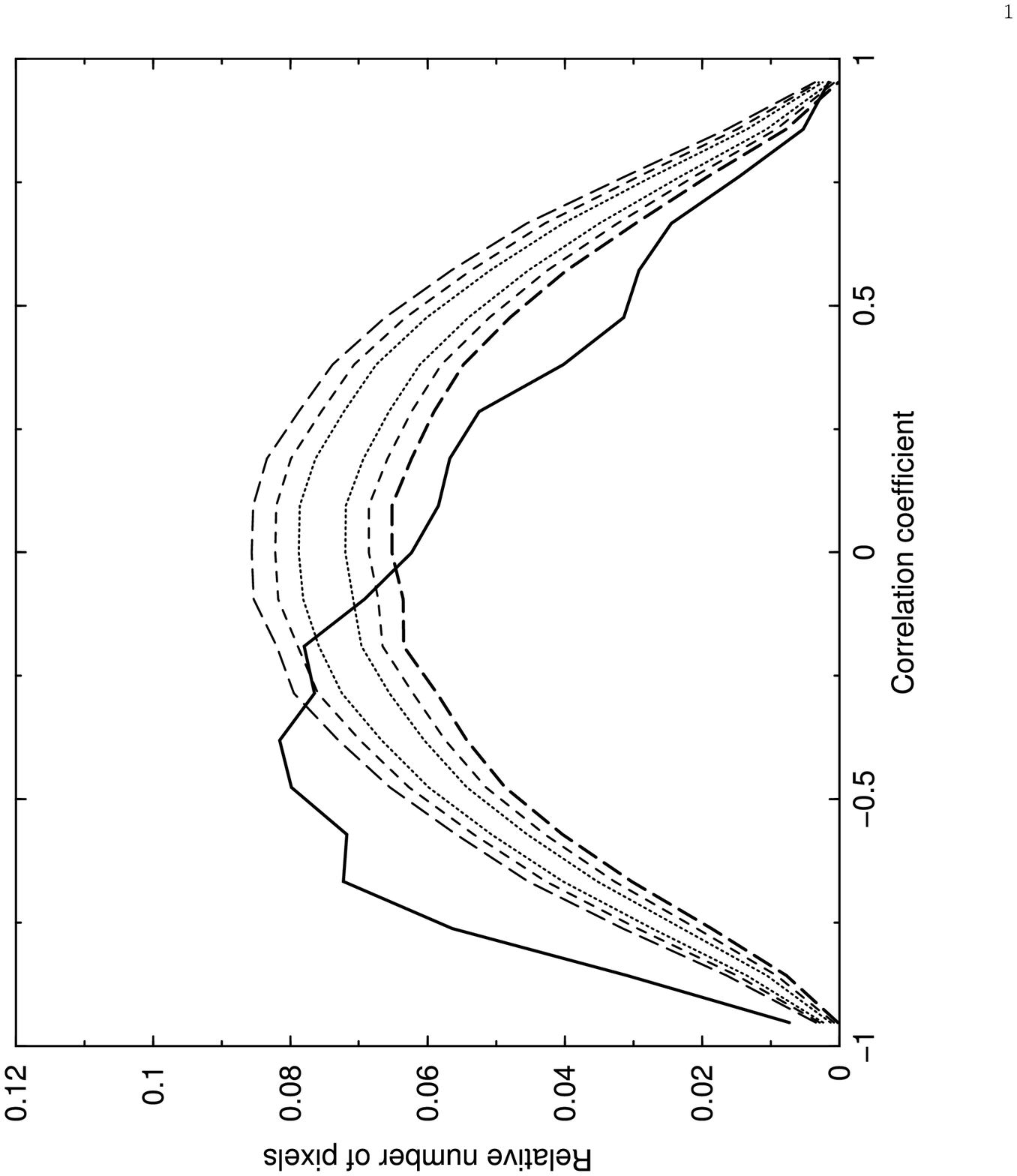,angle=-90,width=9cm}
\psfig{figure=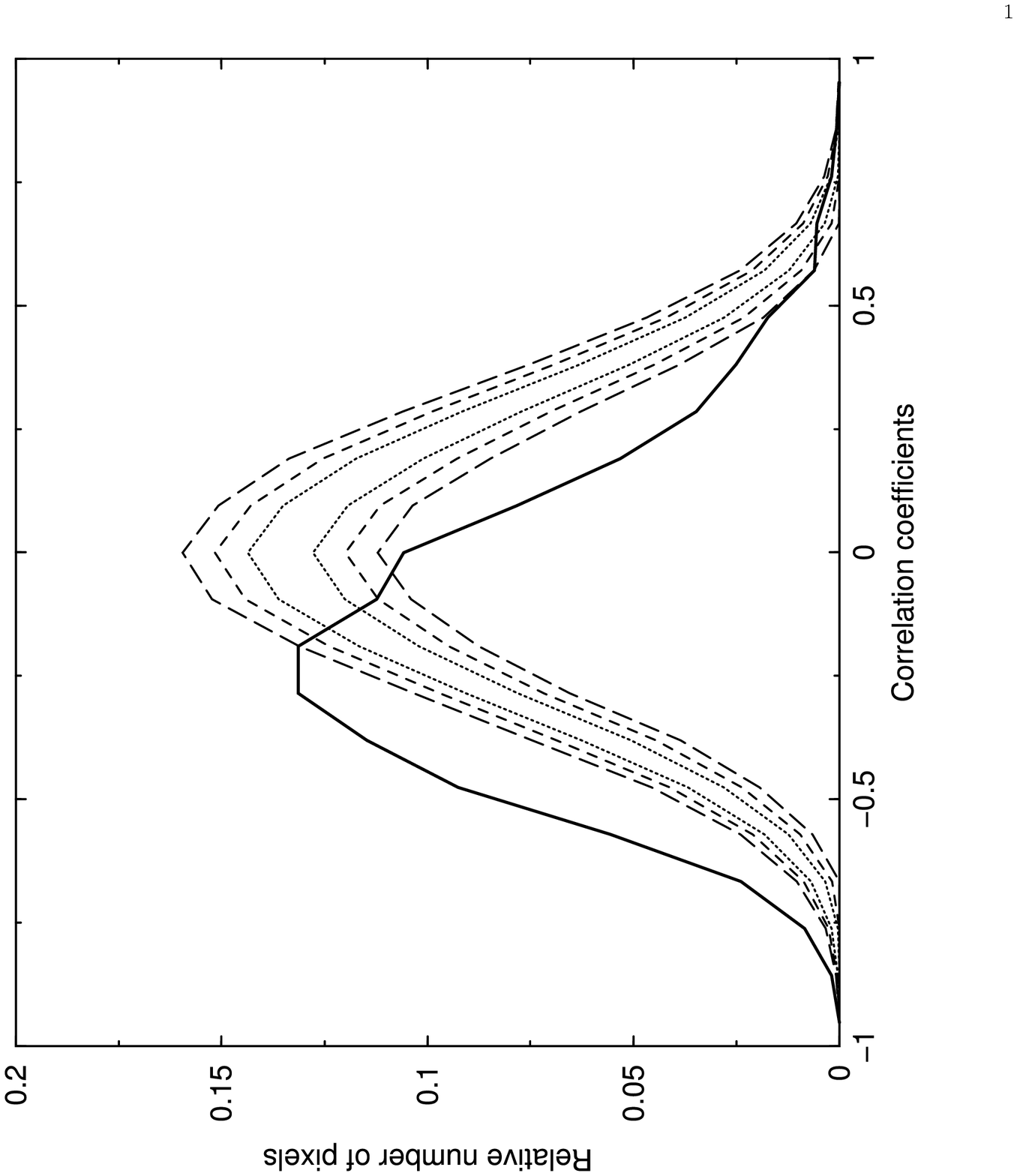,angle=-90,width=9cm}
\psfig{figure=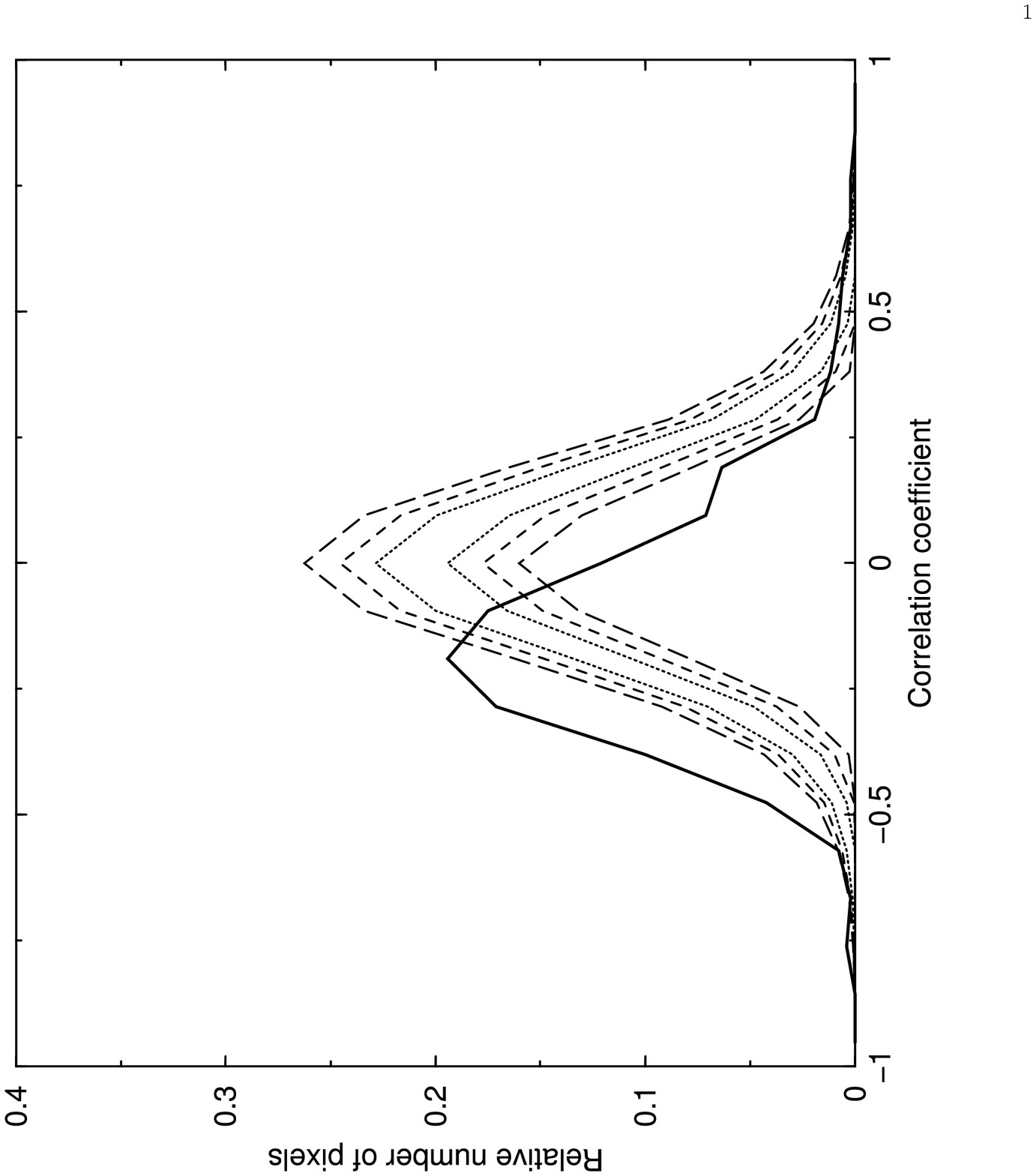,angle=-90,width=9cm}
}} \caption{Histograms of correlation coefficient distribution for
the ILC and dust maps. Top to bottom: histograms for the
pixelization with the pixel side of 160, 300 and 540\arcmin. The
dotted line, short and long dashes mark the $\pm \sigma$, $\pm
2\sigma$ and $\pm$3$\sigma$ levels in the pixel distribution for
correlating random signals, computed for the $\Lambda$CDM
cosmological model and dust.}
\end{figure}

\begin{figure} 
\centerline{\vbox{
\psfig{figure=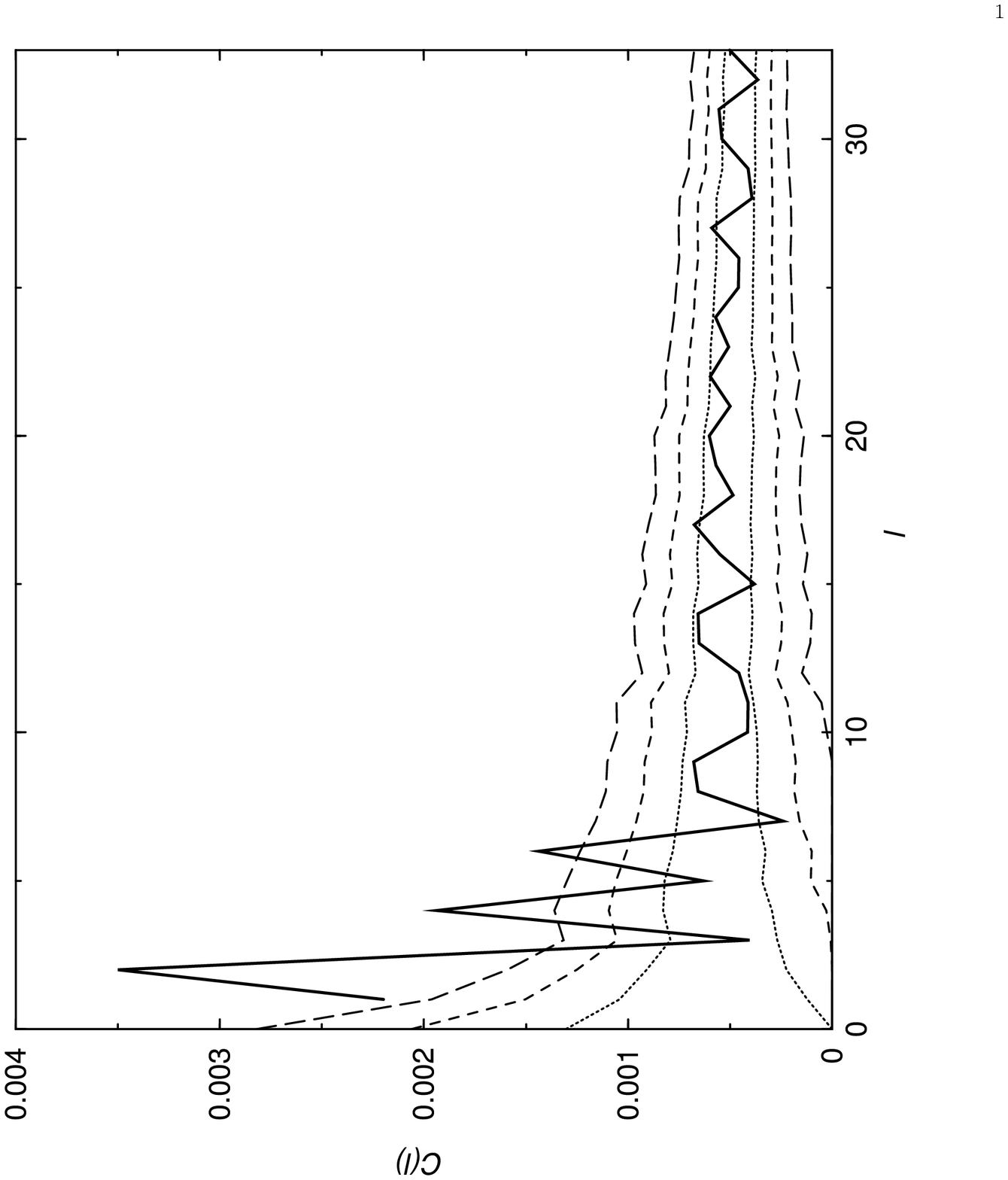,angle=-90,width=8cm}
\psfig{figure=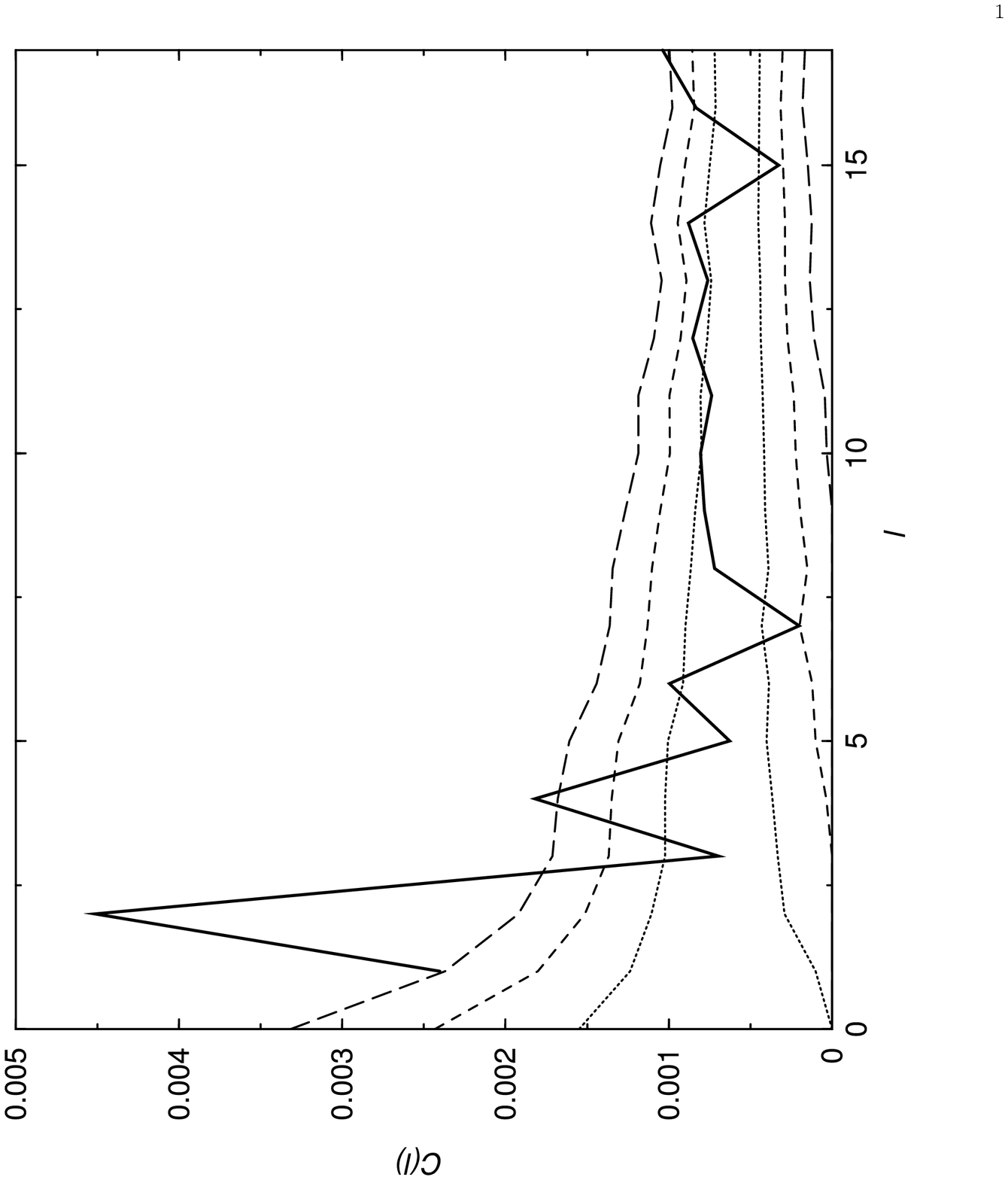,angle=-90,width=8cm}
\psfig{figure=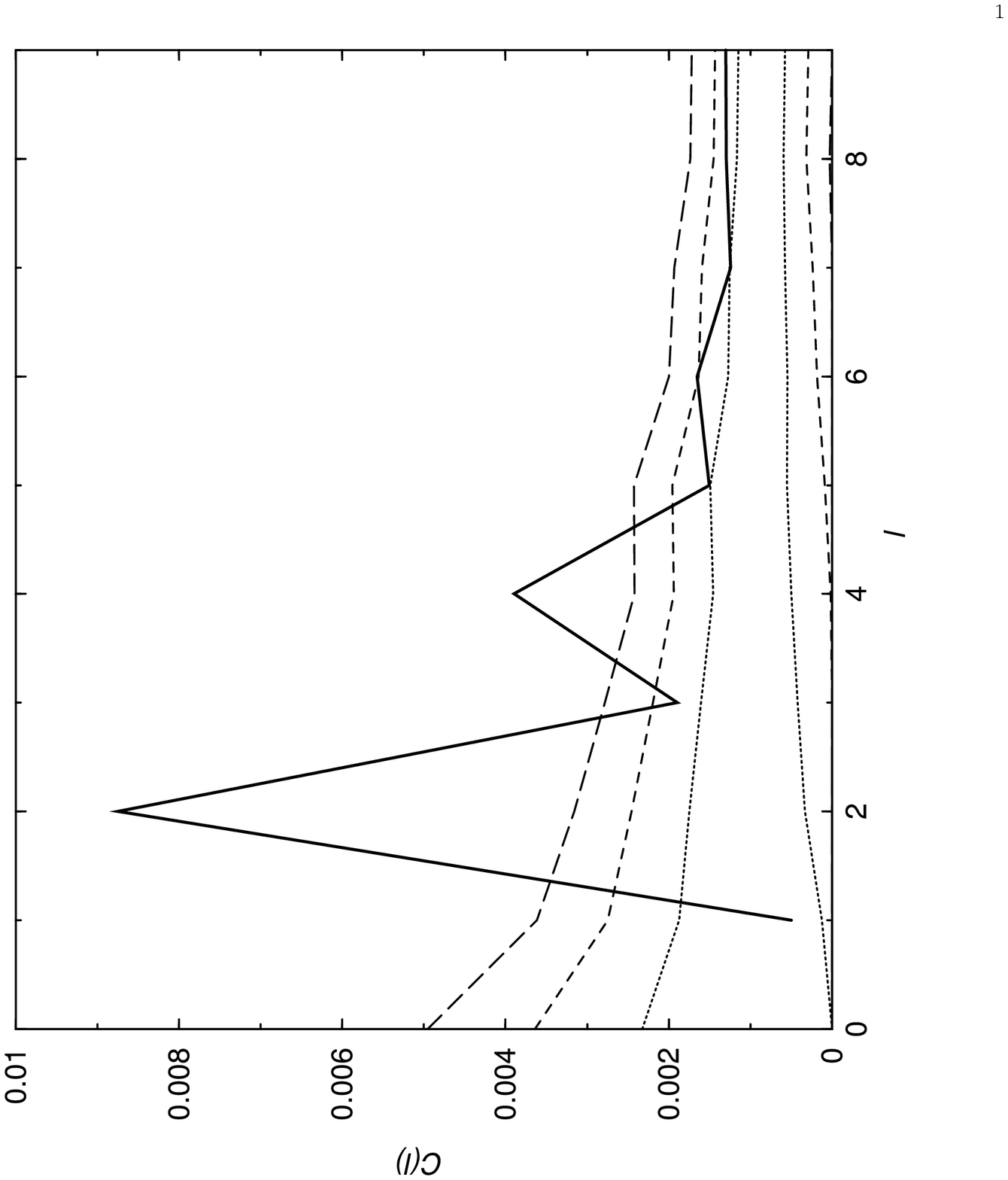,angle=-90,width=8cm}
}} \caption{Power spectra $C(\ell)$ of the correlation coefficient
maps of the ILC signals and dust. Top to bottom: power spectra for
the pixelization with the pixel side of 160, 300 and 540\arcmin.
The dotted line, short and long dashes mark the $\pm \sigma$, $\pm
2\sigma$ and $\pm$3$\sigma$ levels in the scatter of signal values
for the correlation of a random map and dust.}
\end{figure}

The distribution of pixels (correlation coefficients) for the maps
in Fig.\,1 is demonstrated in Fig.\,2. Here the $\pm\sigma$, $\pm
2\sigma$ and \mbox{$\pm$3$\sigma$} levels in the pixel
distribution for the correlation of a random signal and dust are
marked with dotted, short and long dashed lines, respectively.
These levels were computed by averaging the results of
calculations for the 200 random Gaussian maps modelled in terms of
the $\Lambda$CDM cosmology power spectrum. Apart from the shift in
the distribution to the negative values, there is a significant
distortion of the distribution shape, which distinguishes it from
a normal distribution. The median values of the distributions for
the correlation	 pixels of 160\arcmin, 300\arcmin\, and
540\arcmin\, are equal to $-$0.2187, $-$0.2326 and $-$0.2736,
respectively. It should be noted though that the resulting
distribution bias is similar to the bias, obtained at the
statistical evaluations of the quadrupole signal restoration
quality using the ILC method ~\cite{instab:Verkh_ecl_en}.

The shift of the distribution in the negative direction defines
the monopole in the correlation map. It is most likely related to
the procedure of component separation  and is also observed in the
correlations of one-dimensional cross sections
\cite{cr1wmap:Verkh_ecl_en,cr1wm_piter:Verkh_ecl_en}.  However, the distortion may be caused
by the presence	 of enhanced  correlations of higher multipoles in
the maps. To investigate this phenomenon let us construct the
angular power spectrum:
$$
C(\ell)=\frac{1}{2\ell+1}\left[|a_{\ell 0}|^2 +
	      2\sum_{m=1}^\ell |a_{\ell m}|^2\right].
\eqno(6)
$$

The power spectra of the maps with the corresponding
$\ell_{max}=33$, 17 and 9 are demonstrated in Fig.\,3. The dotted,
short and long dashed lines here mark the $\sigma$, 2$\sigma$ and
3$\sigma$ levels in the scatter of spectrum values for the
correlation of random maps and dust. The monopole signal has
previously been removed. It is clear from Fig.\,3 that for all the
three power spectra the quadrupole is much stronger than the other
harmonics. Using the expansion by the formula (1), we extract this
harmonic from the correlation map data. The results are
demonstrated in Fig.\,4.
\begin{figure} 
\centerline{\vbox{
\psfig{figure=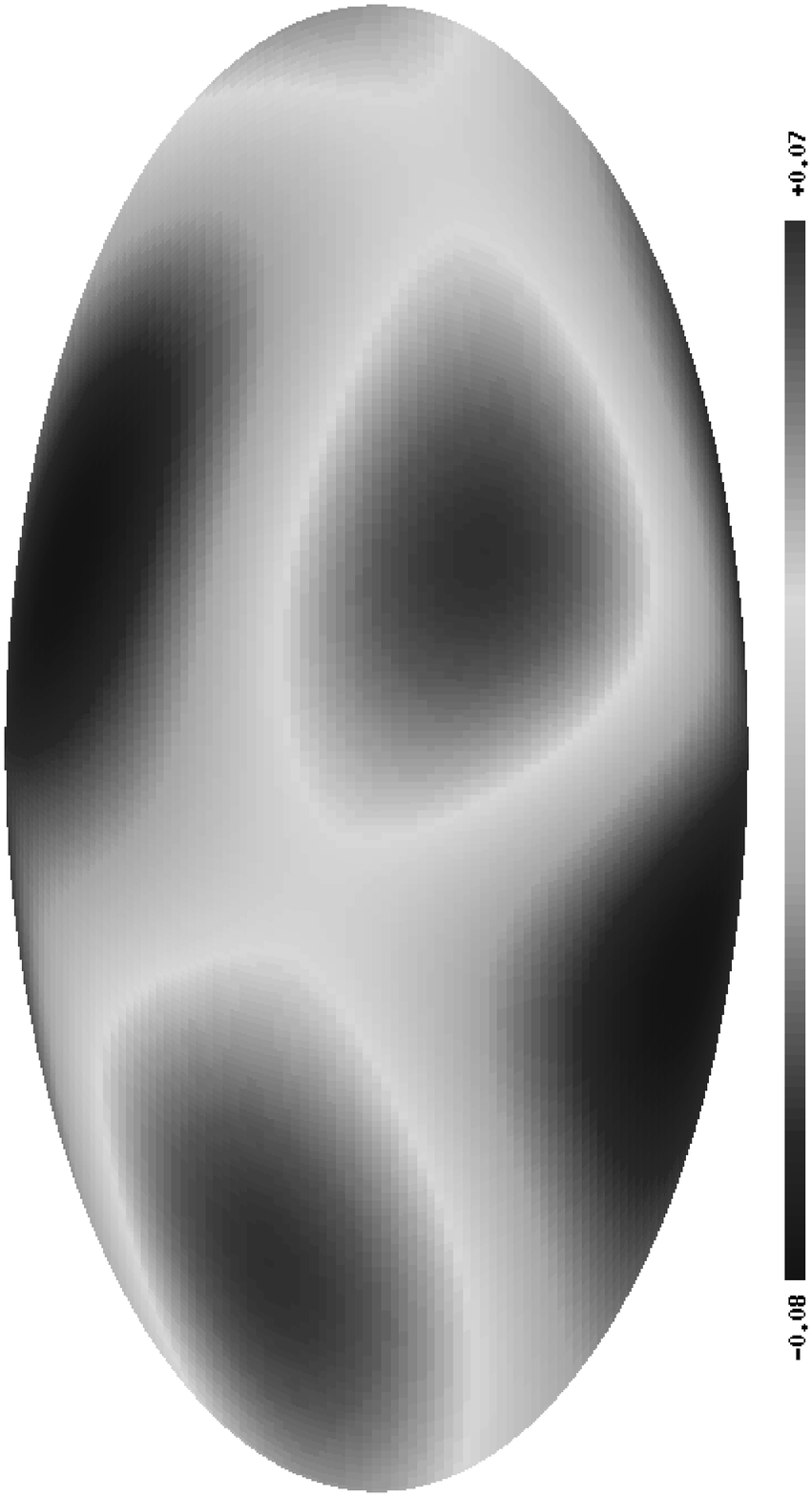,angle=-90,width=9cm}
\psfig{figure=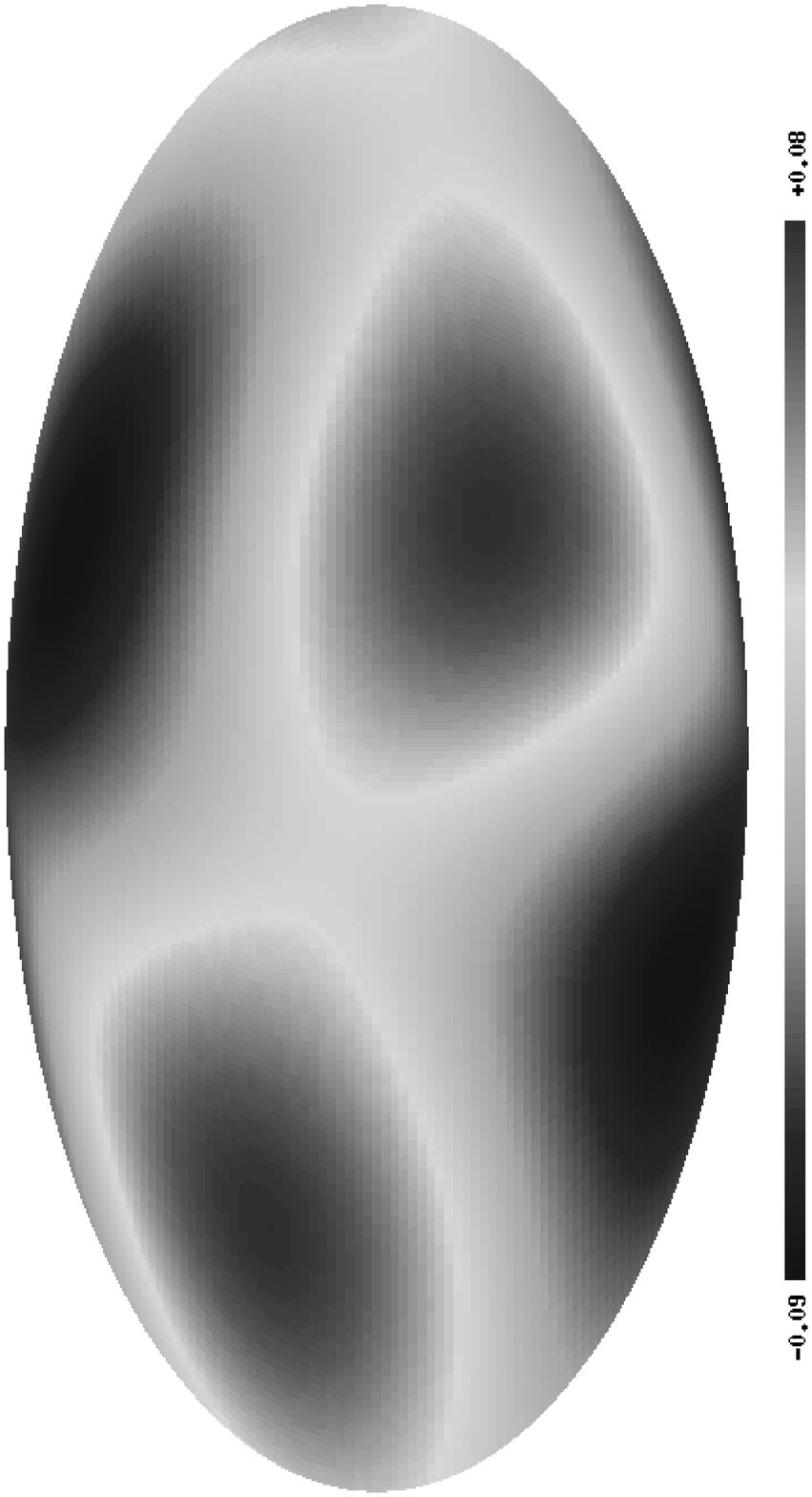,angle=-90,width=9cm}
\psfig{figure=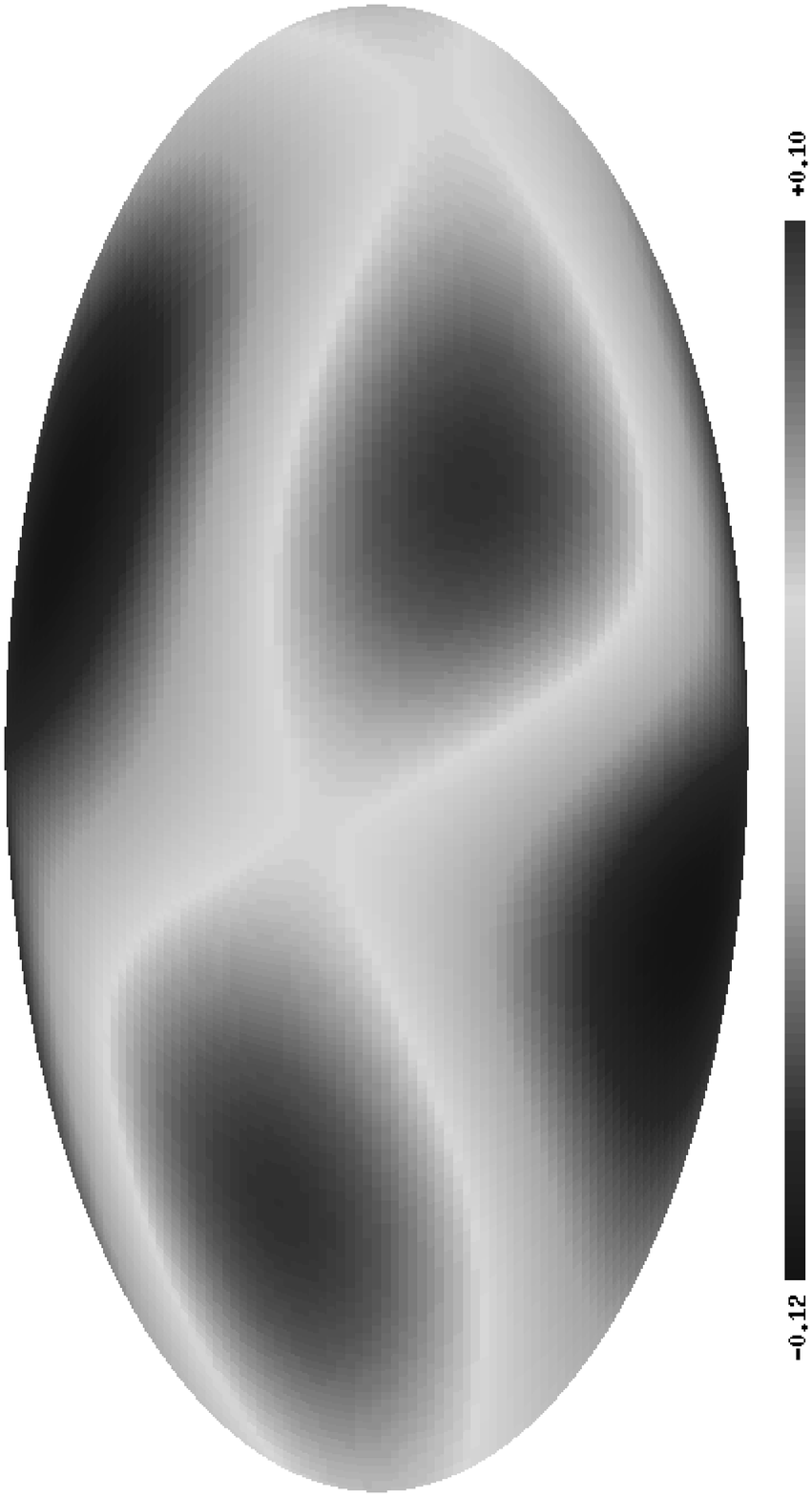,angle=-90,width=9cm}
}} \caption{The quadrupole extracted from the ILC and dust
correlation maps.  Top to bottom: maps for the pixelization of
correlations within the areas with the sides of 160, 300 and
540\arcmin.}
\end{figure}

A peculiarity of this quadrupole is its sensitivity to both
coordinate grids we selected: ecliptic and equatorial, as
demonstrated in Fig.\,5. The figure shows that the cold spots near
the galactic poles lie in the ecliptic plane, and hot spots lie
exactly in the equatorial poles.

\begin{figure*} 
\centerline{\hbox{
\psfig{figure=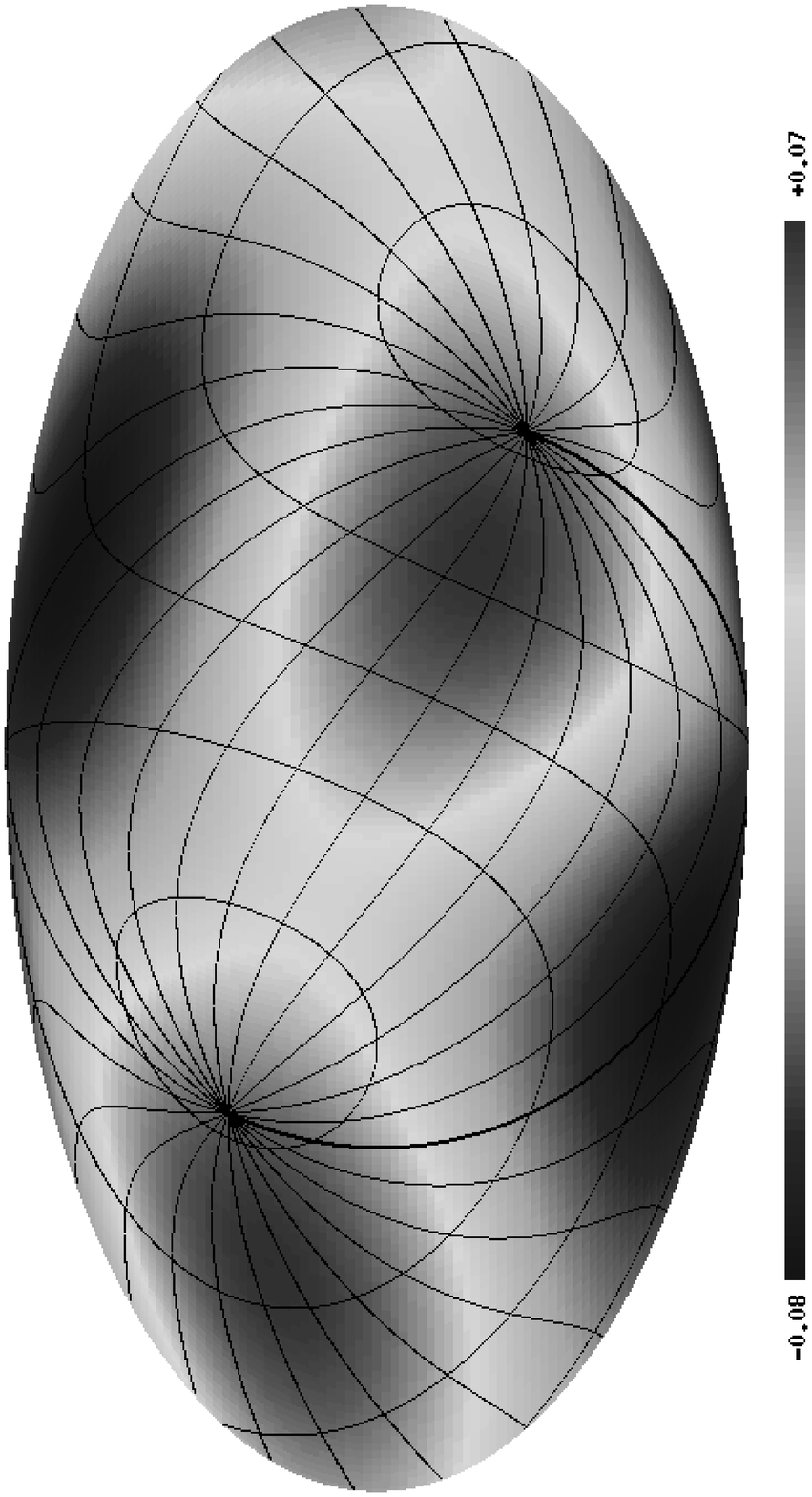,angle=-90,width=7cm}
\psfig{figure=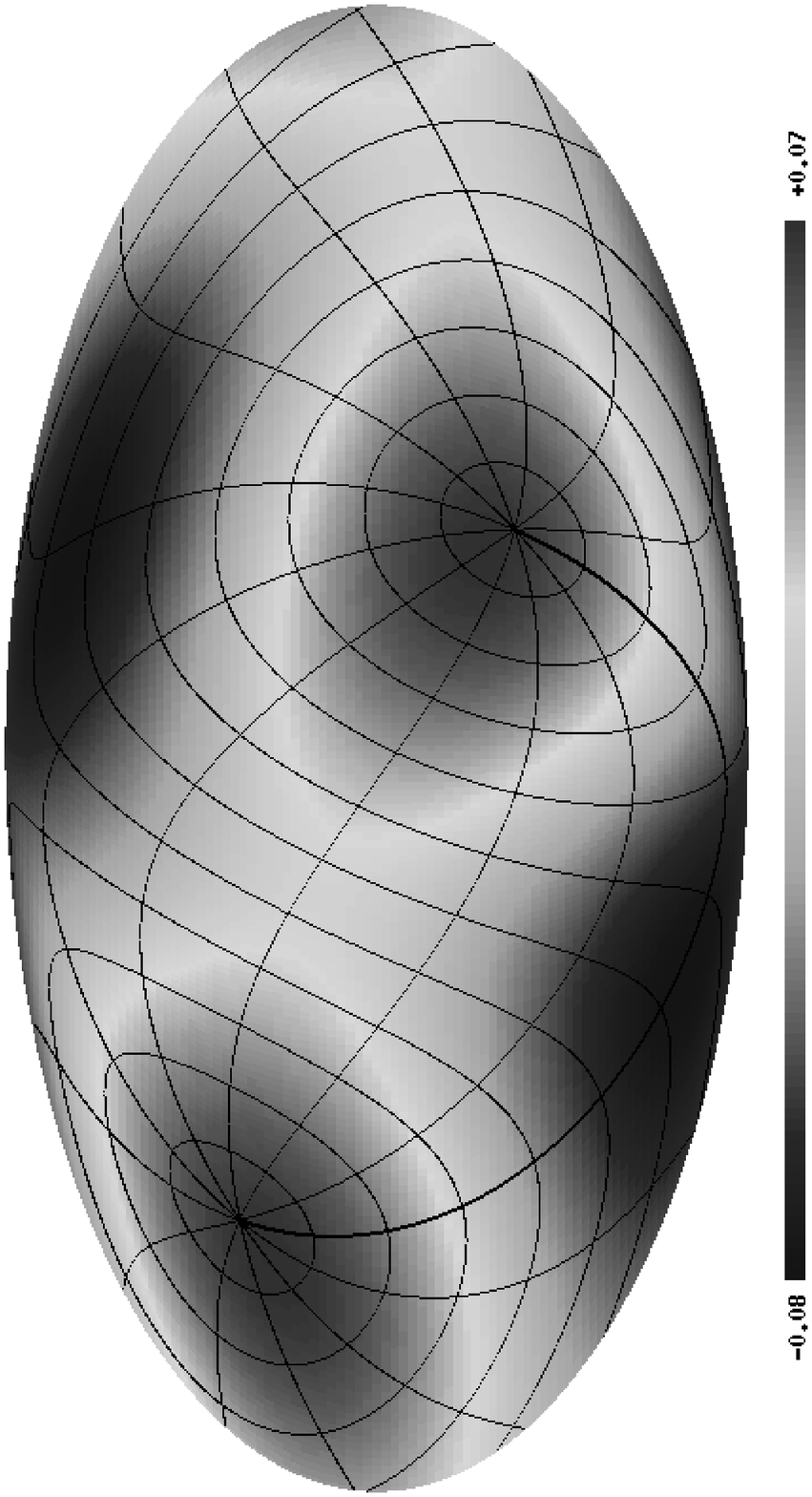,angle=-90,width=7cm}
}} \caption{Ecliptic (left) and equatorial (right) coordinate
grids in the quadrupole map, constructed based on the data with
the  pixelization of 160\arcmin. The coordinate grid poles fall on
the hot spots (light gray), and the cold spots (dark gray) lie in
the equator. }
\end{figure*}

Note as well the fourth harmonic in the power spectrum for these
correlations with the scale of 160\arcmin. Its amplitude is also
higher than the	 3$\sigma$ level. It is evident from the map of
this multipole (Fig.\,6) that this correlation is determined by
the structures in the Galactic plane, which, in general, is
anticipated on the scales of 15\degr--20\degr.
\begin{figure} 
\centerline{\hbox{
\psfig{figure=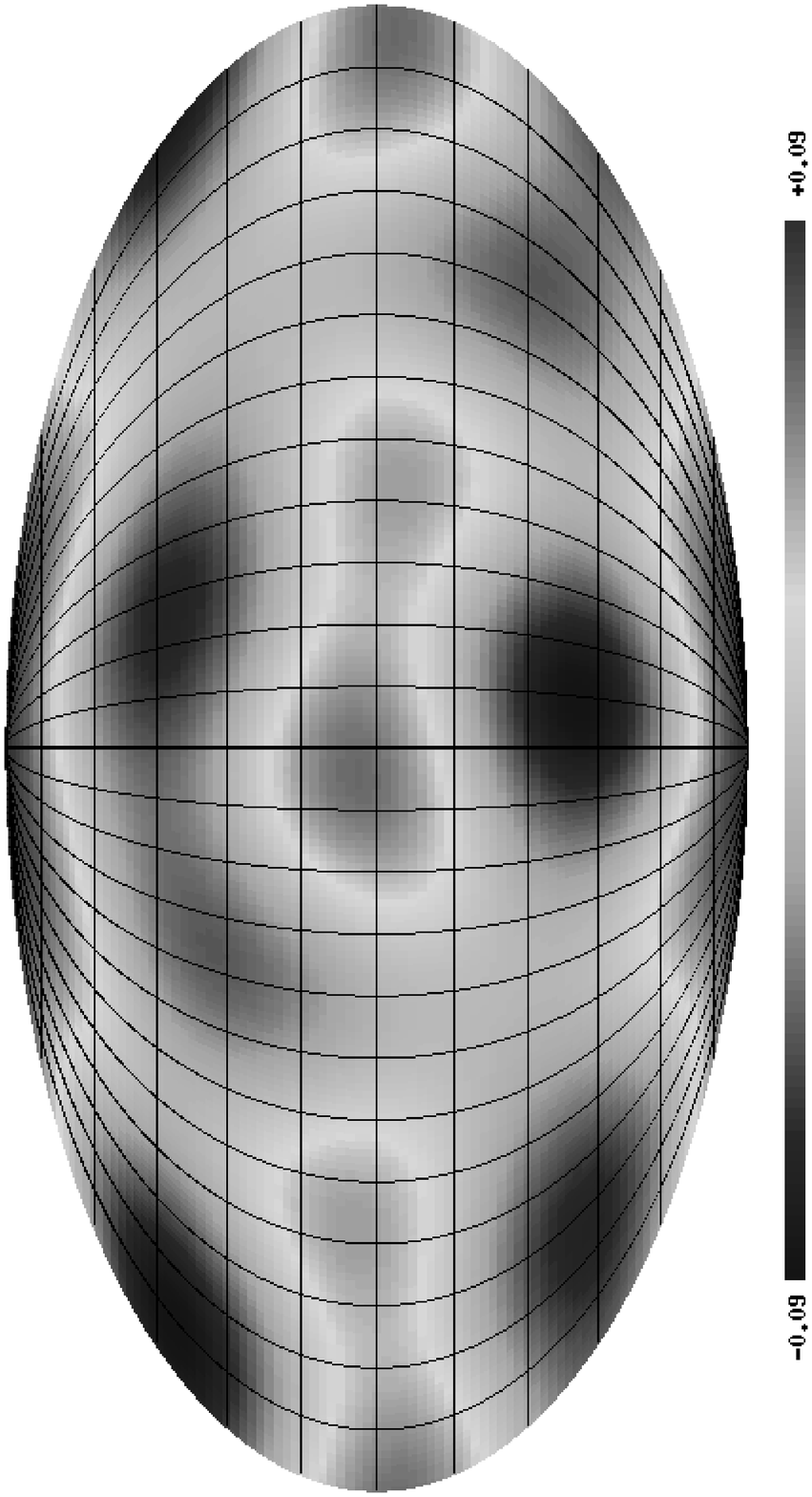,angle=-90,width=7cm}
}} \caption{Map of the fourth multipole with a superimposed
galactic coordinate grid. The map is calculated for the
correlation window with the resolution of  160\arcmin.}
\end{figure}

Returning to the quadrupole, we show that the removal of the
Galactic plane from the calculations does not affect the position
of this harmonic. For this purpose, we use the Kp0 mask, which is
available from the  WMAP mission website\footnote{\tt
http://lambda.gsfc.nasa.gov} and is applied for shielding the
Galactic plane and point source radiation.

\begin{figure*} 
\centerline{\vbox{
\hbox{
\psfig{figure=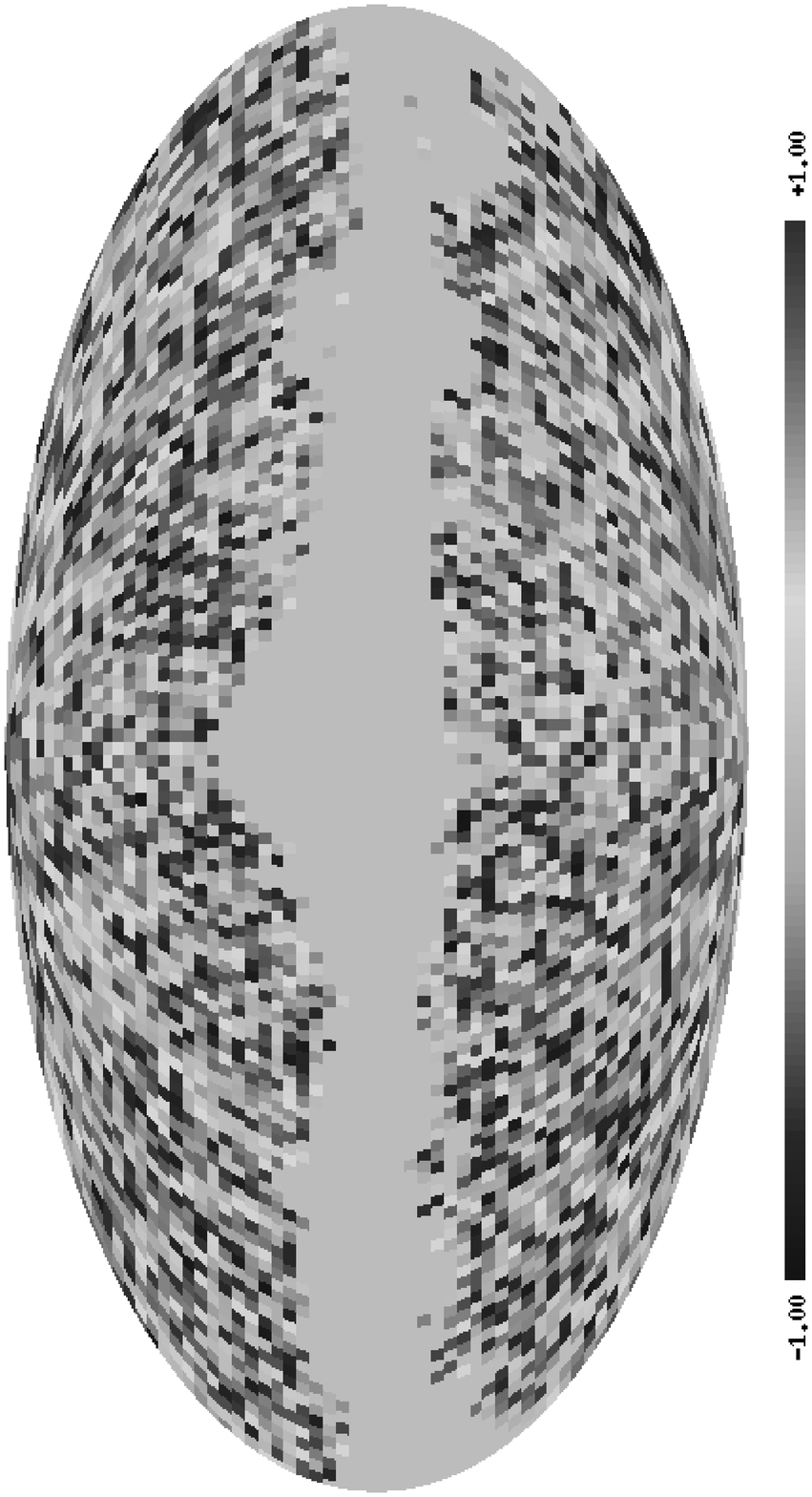,angle=-90,width=6cm}
\psfig{figure=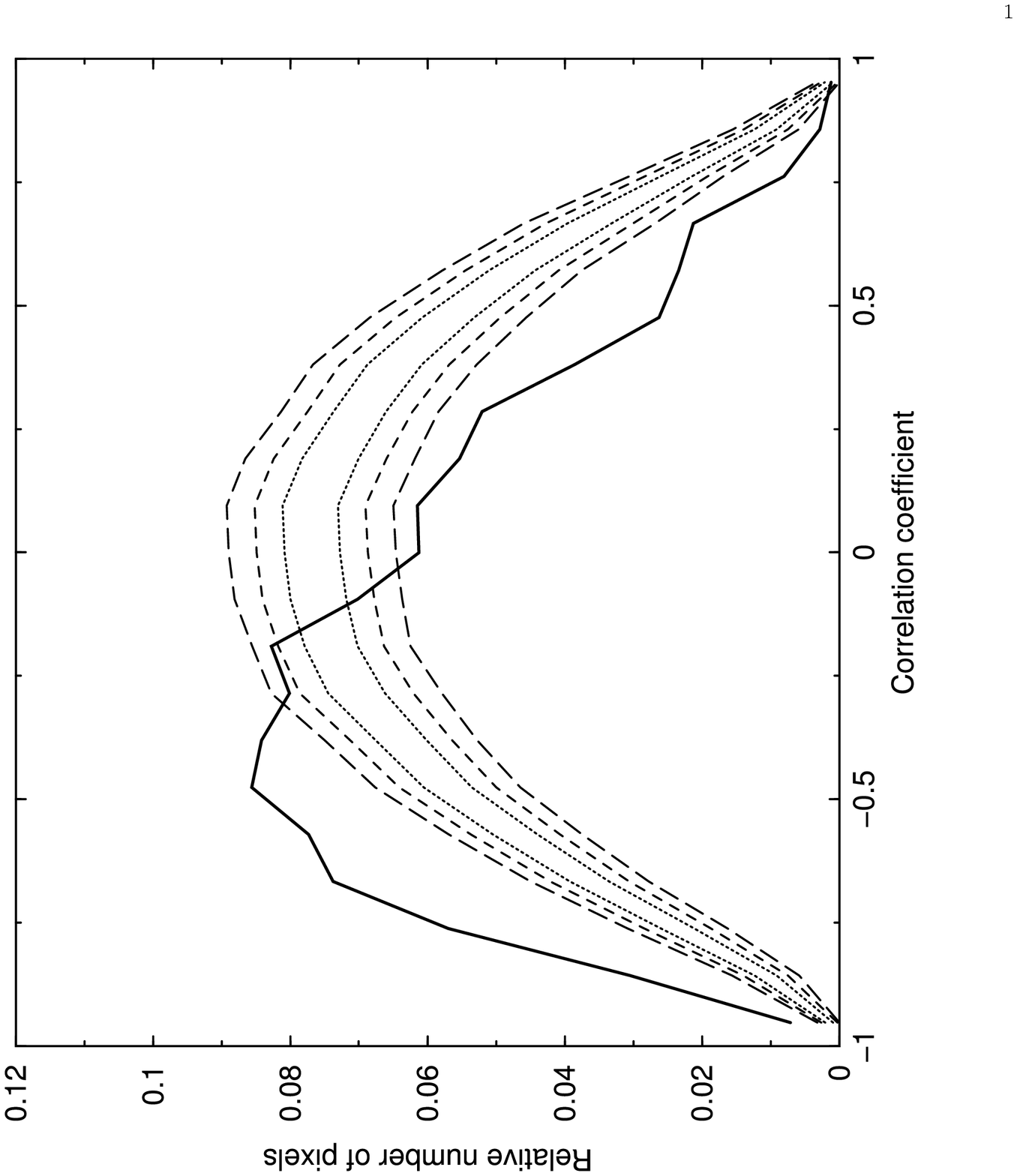,angle=-90,width=5cm}
} \hbox{
\psfig{figure=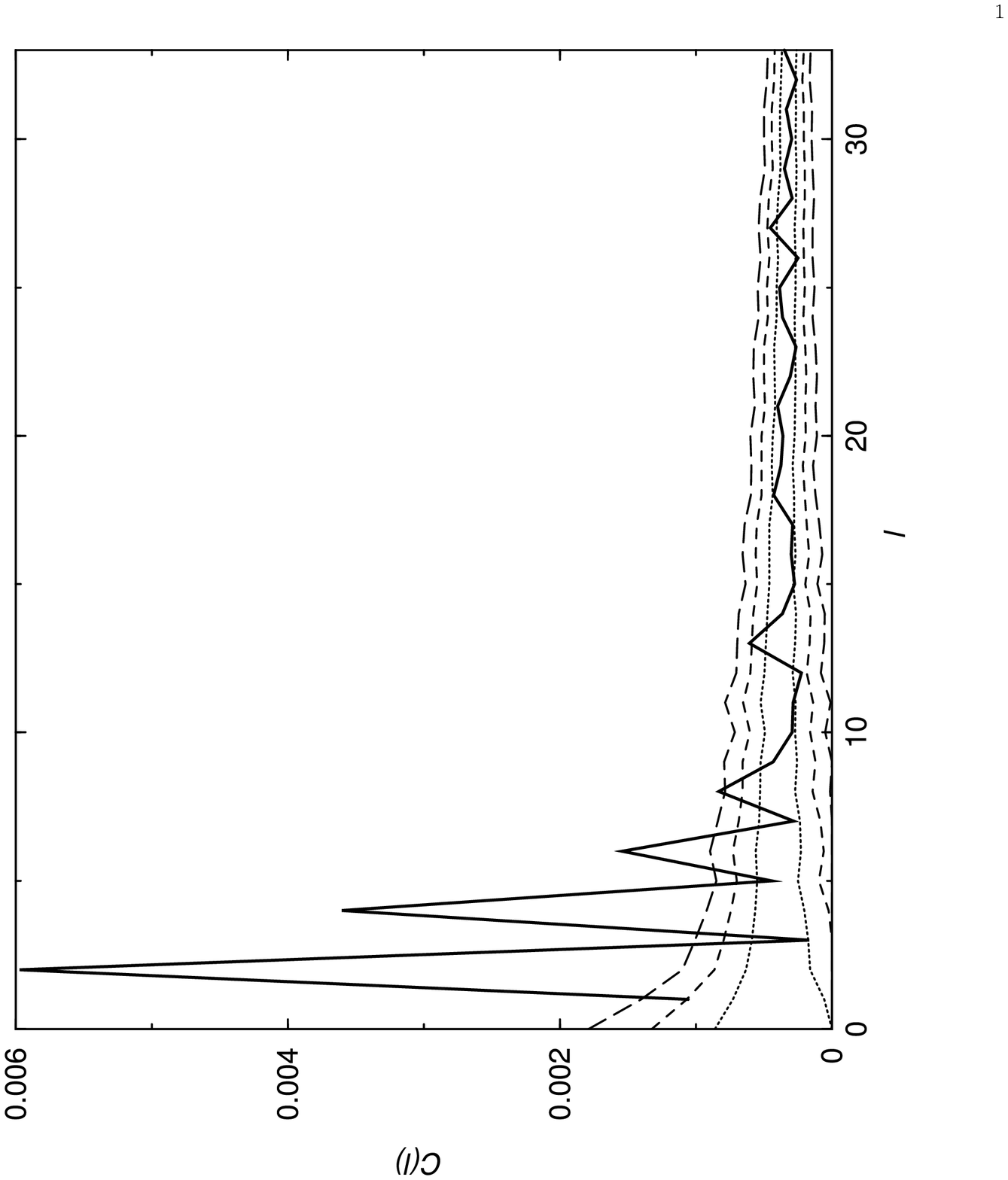,angle=-90,width=5cm}
\psfig{figure=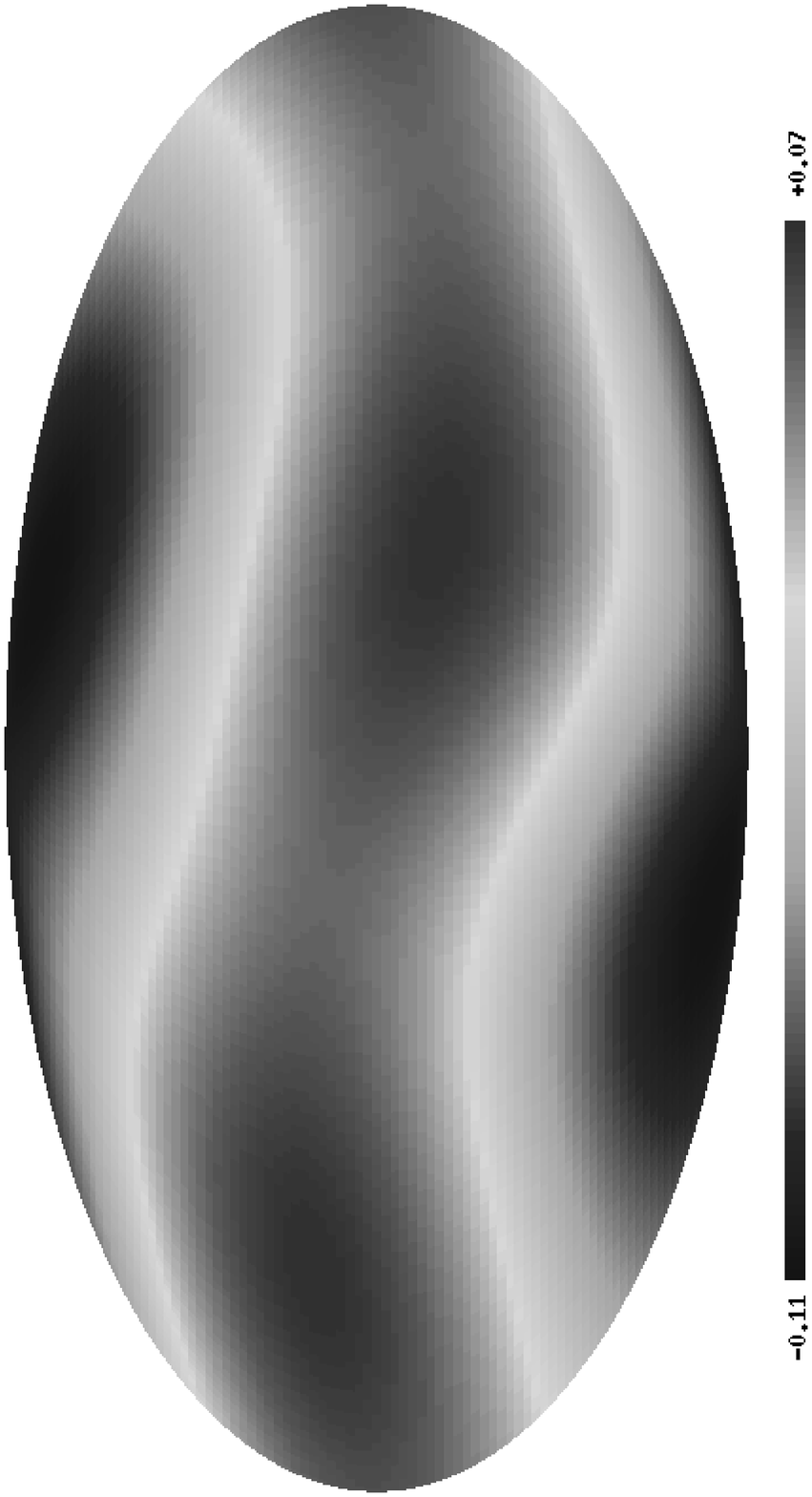,angle=-90,width=6cm}
}}} \caption{ILC and dust signal correlation map for the
160\arcmin\, correlation window	 and its statistical diagrams. The
Kp0 mask  is superimposed on the map. The diagrams were calculated
accounting for the mask. Top left: a correlation map for the pixel
side of 160\arcmin. Top right: a histogram of pixel distribution
in this map outside the Kp0 mask. The dotted, short and long
dashed lines mark the $\sigma$, 2$\sigma$ and 3$\sigma$ levels in
the scatter of the signal values for the correlation of a random
map and dust with a superimposed mask. Bottom left: the power
spectrum. Dotted and dashed lines demonstrate the scatter,
estimated similarly to the previous figure. Bottom right: the
quadrupole map.}
\end{figure*}

Figure\,7 shows the map of dust and ILC signal correlations in the
areas with a side of 160\arcmin\, with a superimposed Kp0 mask. It
also shows a histogram of counts outside the mask. Although a full
power spectrum has no physical meaning in this case, we still use
it to assess the statistical properties of the signal, applying a
set of simulated maps with a superimposed mask. A statistical
analysis can be then done based on the comparative
characteristics. Figure\,7 shows the angular power spectrum
calculated on the full sphere, where the pixels located in the Kp0
mask area have the zero values inscribed. The dotted and dashed
lines show the $\sigma$, 2$\sigma$ and 3$\sigma$ levels	 in the
scatter of the signal values for the correlation of a random map
and dust with a superimposed mask. Note that the ratio of the
quadrupole amplitude in the spectrum in the studied correlation
map  to the 3$\sigma$ level is close to that anticipated from
random maps, without the use of the mask it is equal to 2.2, while
in the case of mask this ratio amounts to 5.5. Therefore shielding
the Galactic plane increases the significance of the quadrupole
harmonic contribution in the correlation map. Figure\,7 as well
demonstrates the quadrupole map, the axis of maxima of which is
turned with respect to that shown in Fig.\,4 due to the presence
of the mask in the spectral analysis, and the axis of minima is
unchanged.

\section{CORRELATIONS BETWEEN ILC AND SYNCHROTRON BACKGROUND COMPONENT MAPS}

In \cite{cormap:Verkh_ecl_en} we discussed the ILC and synchrotron radiation
signal correlation maps. These maps are as well visually revealing
the Galactic plane (Fig.\,8), but lack a clear cut deviation of
the histogram from the Gaussian distribution (Fig.\,9), as it is
the case of the ILC and dust emission signal distribution
correlations (Fig.\,2). But note that for all the three count
distribution histograms, constructed for the 160, 300 and 540
\arcmin windows, there still exists a deviation from the Gaussian
case above the 1$\sigma$ level.

\begin{figure} 
\centerline{\vbox{
\psfig{figure=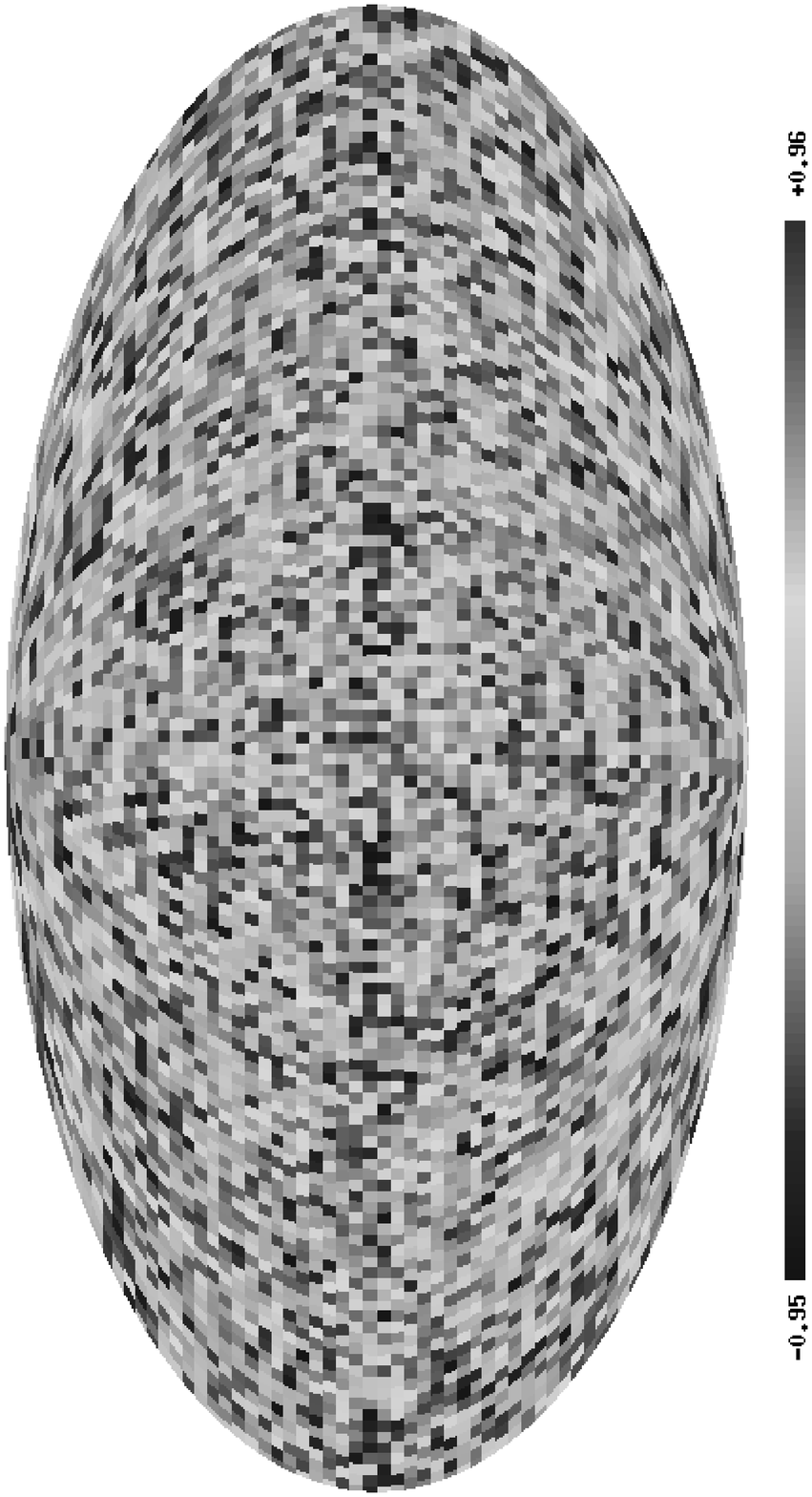,angle=-90,width=9cm}
\psfig{figure=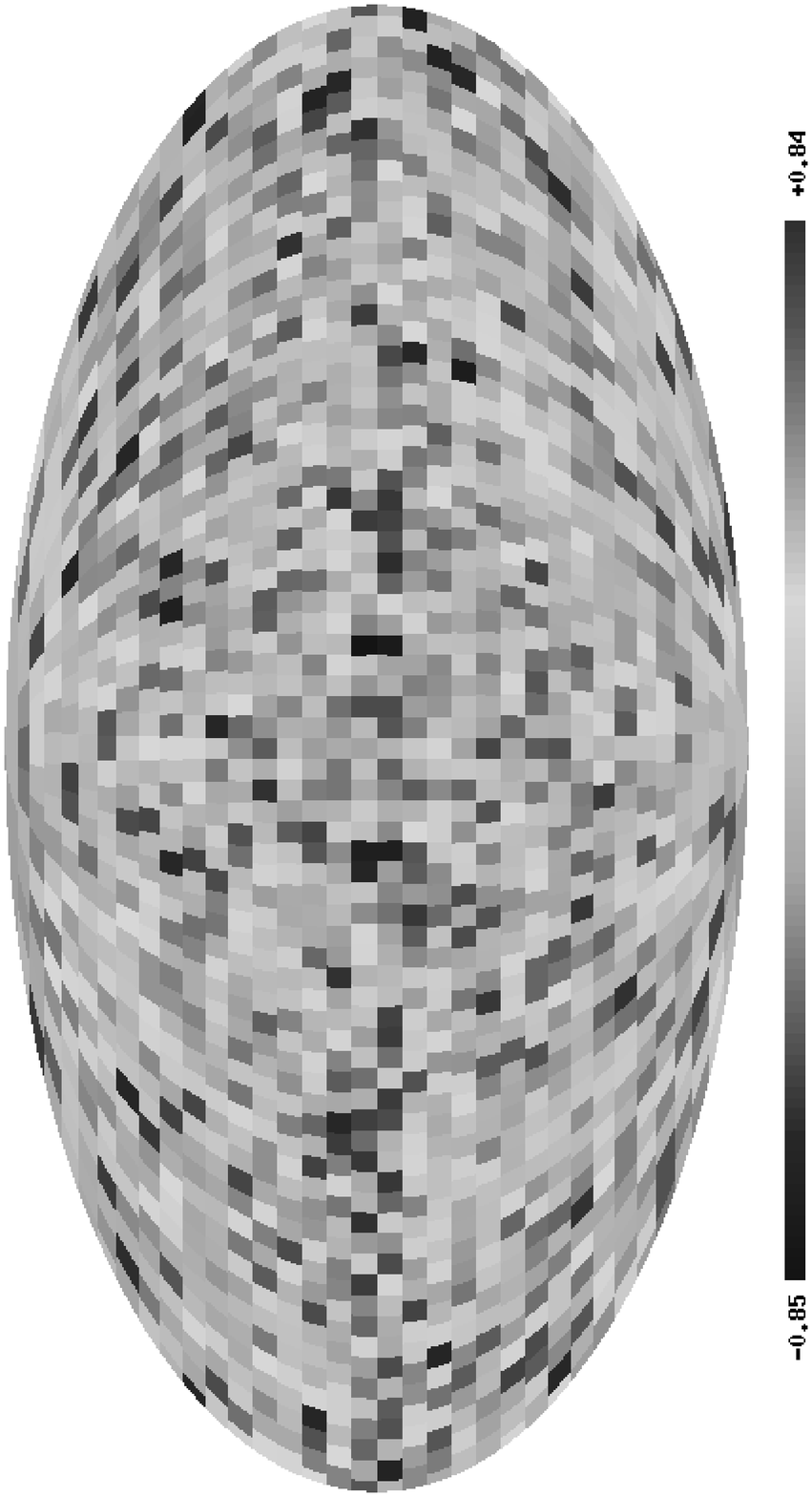,angle=-90,width=9cm}
\psfig{figure=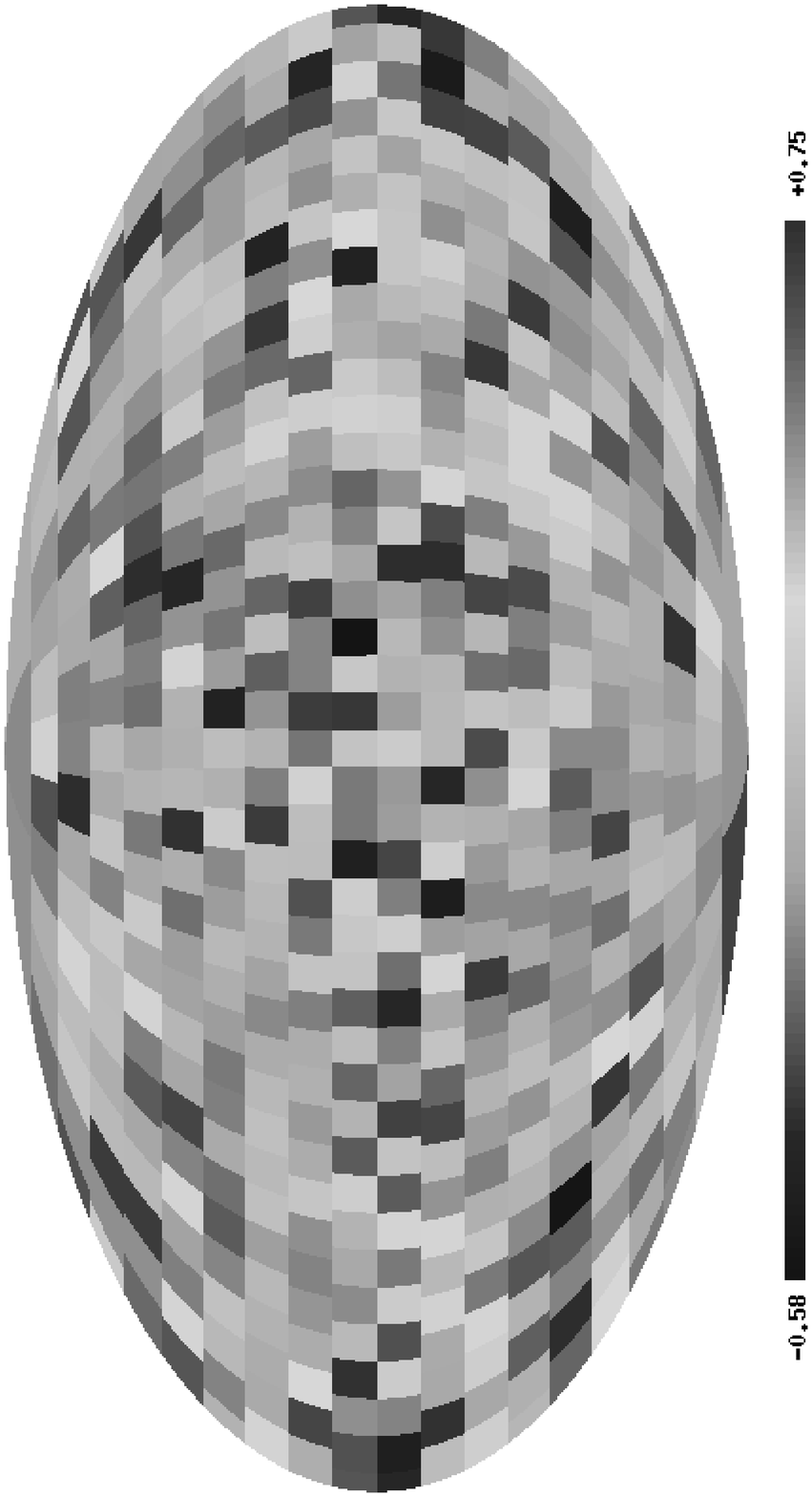,angle=-90,width=9cm}
}} \caption{Maps of correlation coefficients on the sphere,
constructed for mosaic correlations of the ILC and synchrotron
radiation maps. Top to bottom: maps for the pixelization with the
pixel side of 160\arcmin, 300\arcmin and 540\arcmin.}
\end{figure}

\begin{figure} 
\centerline{\vbox{
\psfig{figure=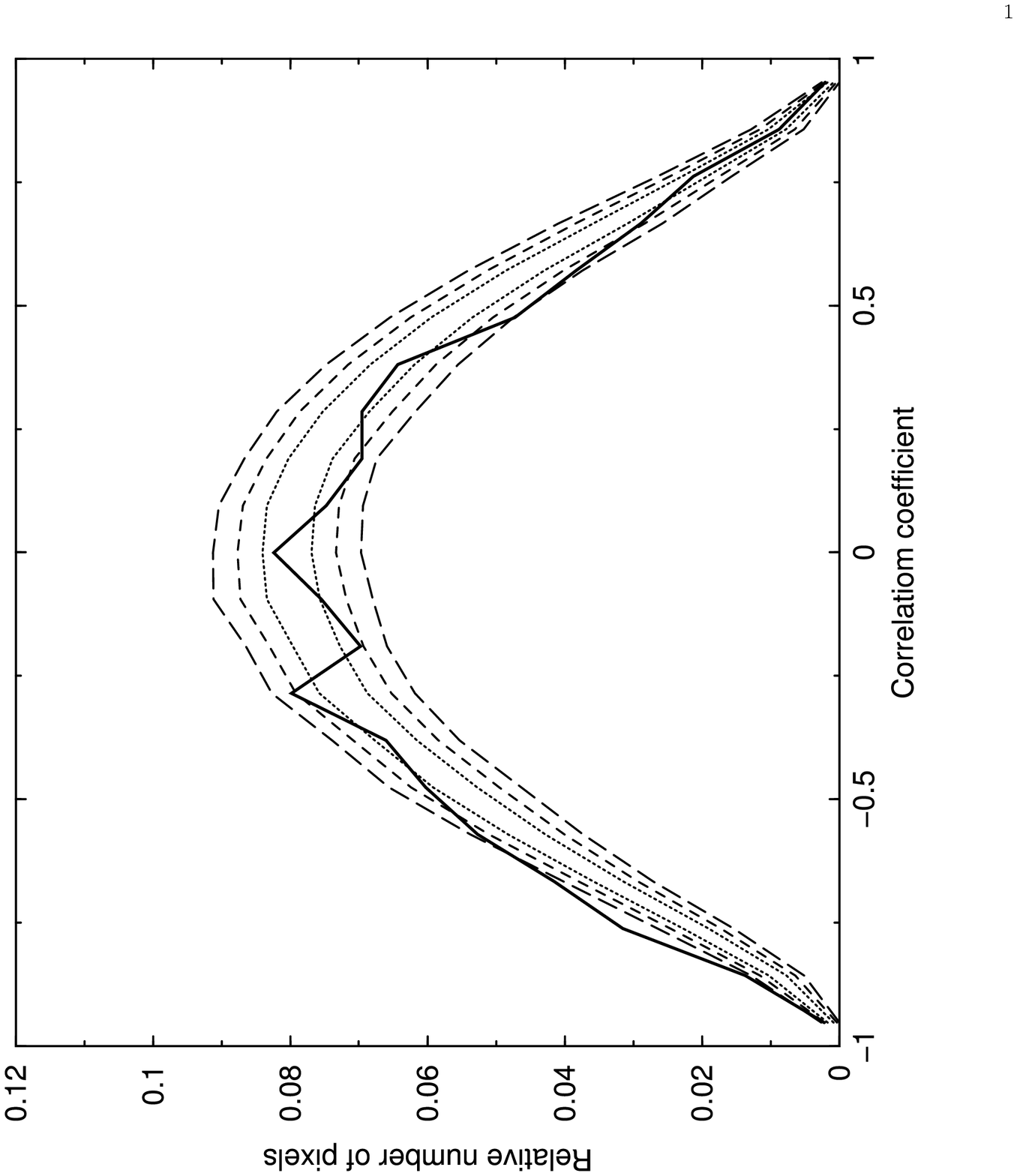,angle=-90,width=8cm}
\psfig{figure=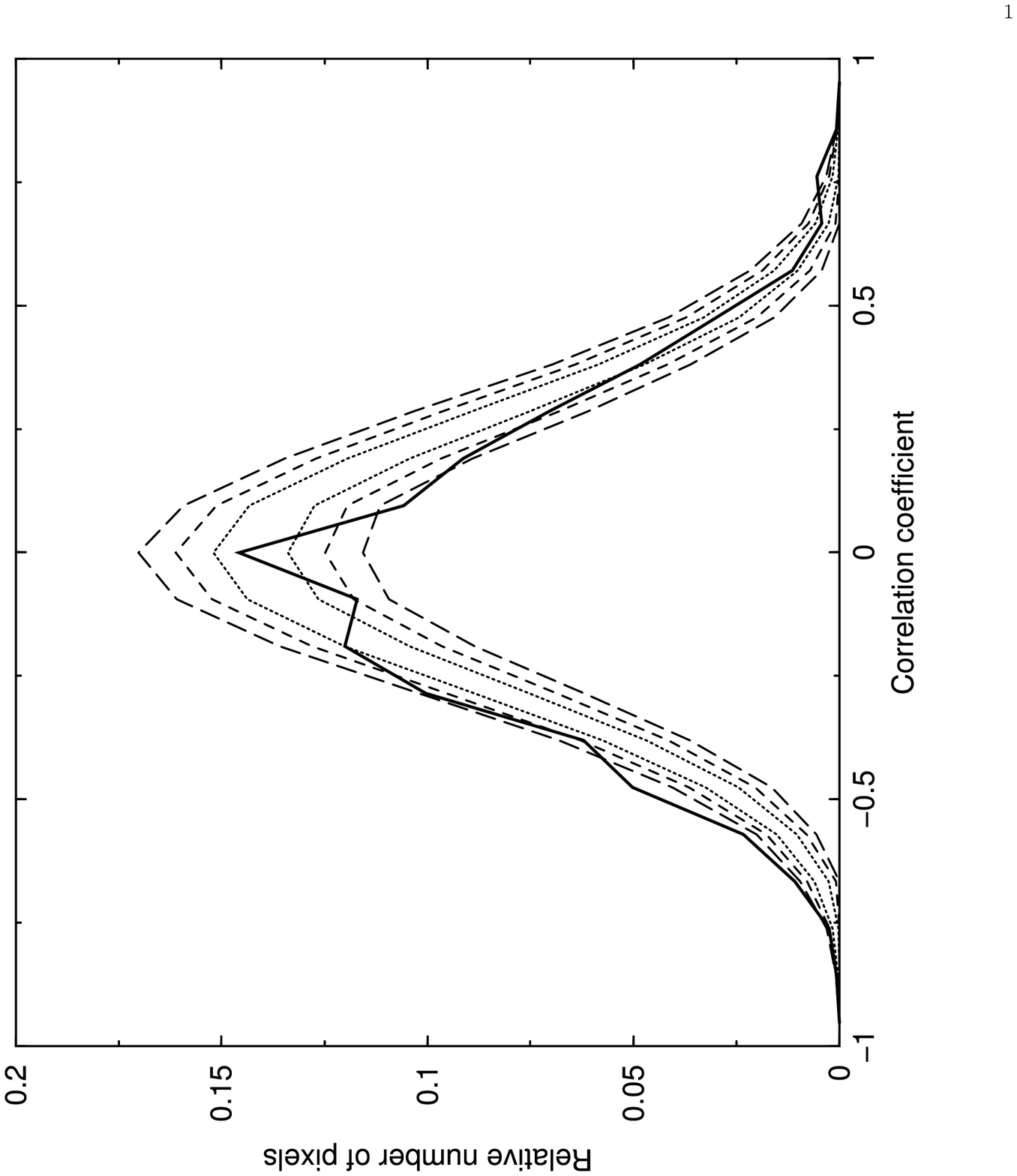,angle=-90,width=8cm}
\psfig{figure=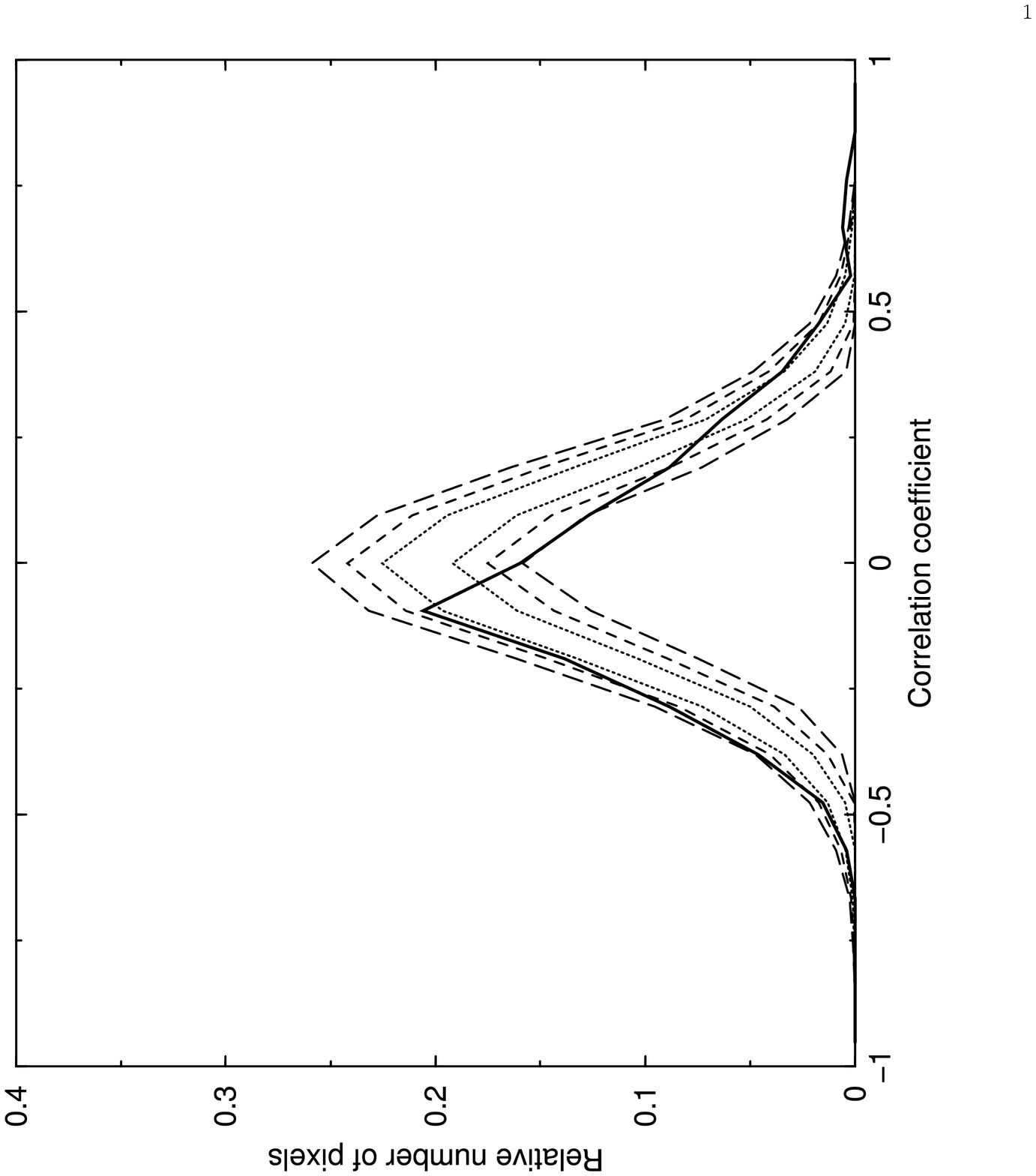,angle=-90,width=8cm}
}} \caption{Histograms of correlation coefficient distribution for
the ILC and synchrotron radiation maps. Top to bottom: histograms
for the pixelization with the pixel side of 160\arcmin, 300\arcmin and
540\arcmin. The dotted line, short and long dashes mark the $\pm
\sigma$, $\pm 2\sigma$ and $\pm$3$\sigma$ levels in the pixel
distribution for correlating random signals and synchrotron
radiation. The random signal is computed for the $\Lambda$CDM
cosmological model.}
\end{figure}

Let us build the power spectra of the mosaic correlation maps for
the ILC signal and synchrotron radiation in the same manner as we
previously did for the dust maps. They are demonstrated in
Fig.\,10 for various correlation windows. Each figure also shows
the limits of the signal scatter at the	 $\pm \sigma$, \mbox{$\pm
2\sigma$} and $\pm$3$\sigma$ levels, calculated from the data of
200 correlation maps for the random signal in the  $\Lambda$CDM
cosmology. We pre-subtracted the monopole from the maps with the
160 and 300\arcmin correlation windows.

\begin{figure} 
\centerline{\vbox{
\psfig{figure=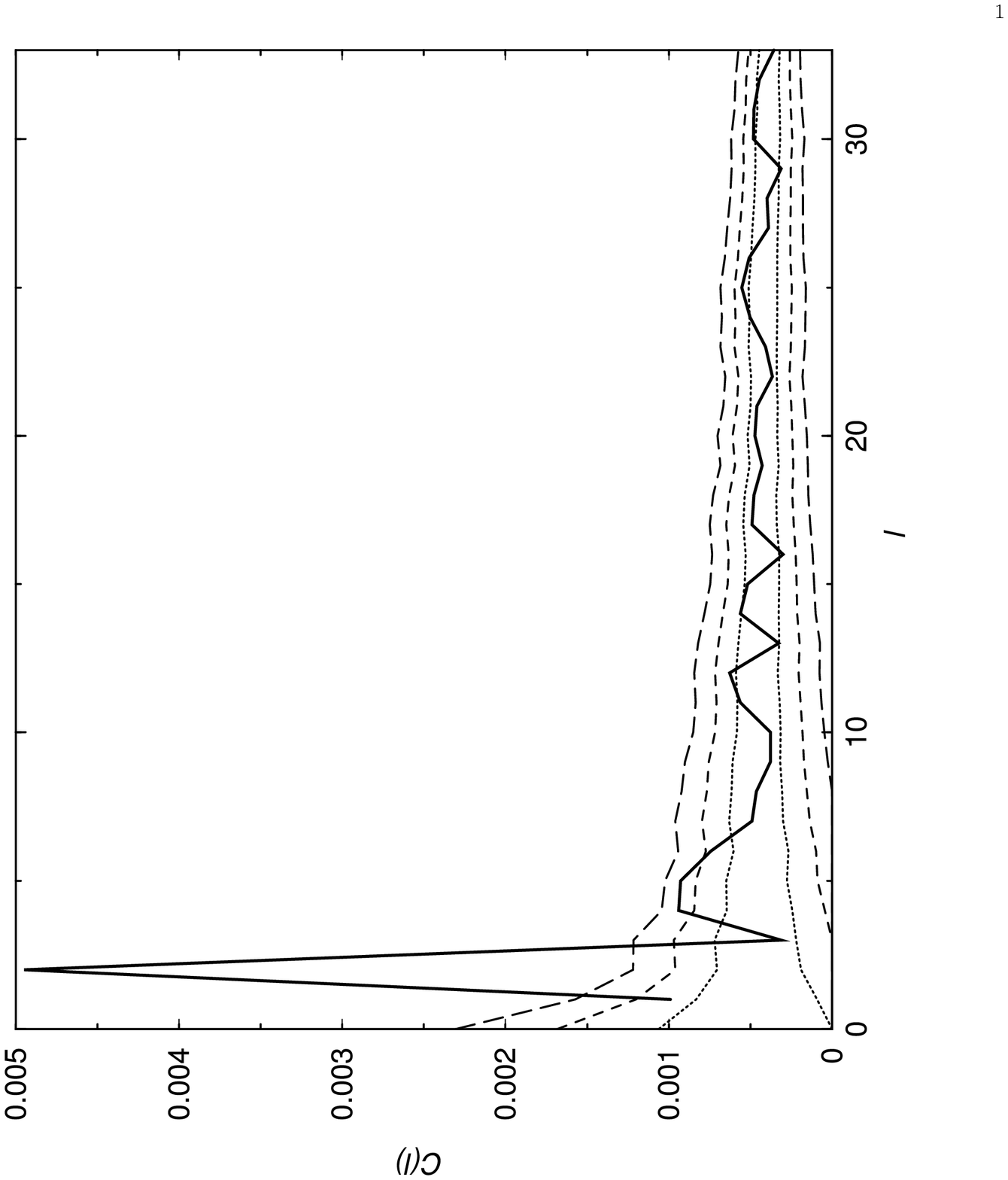,angle=-90,width=8cm}
\psfig{figure=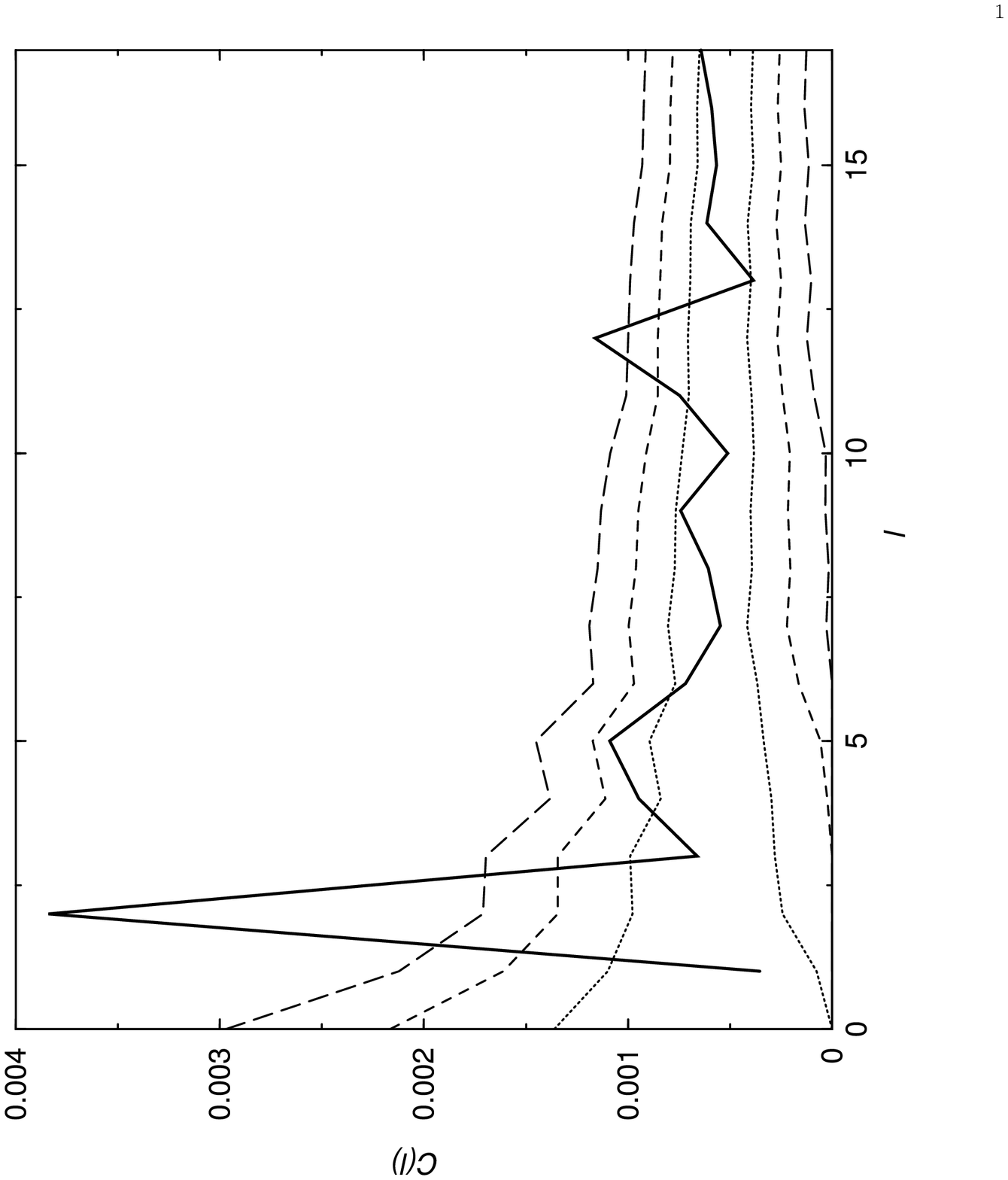,angle=-90,width=8cm}
\psfig{figure=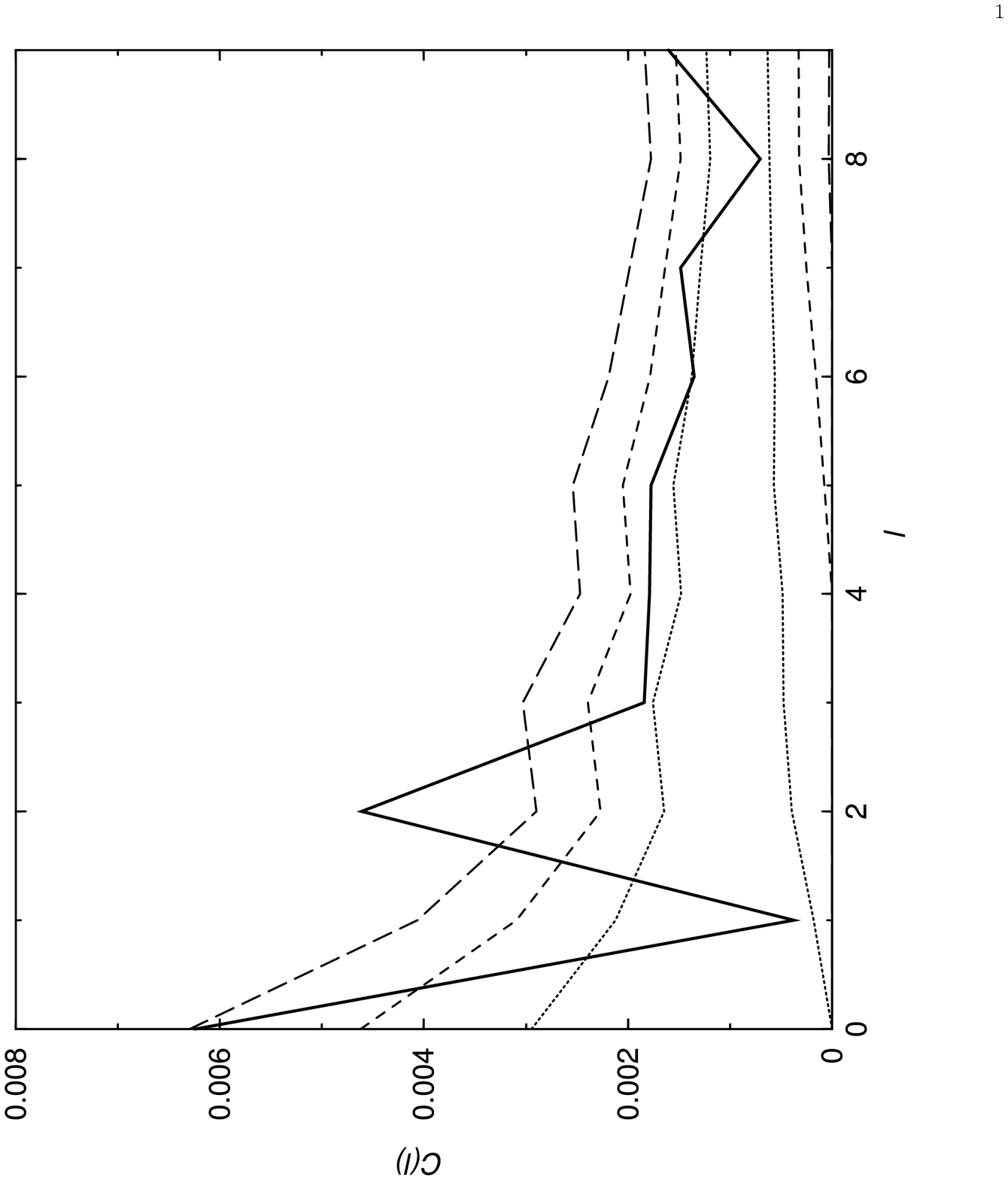,angle=-90,width=8cm}
}} \caption{Power spectra $C(\ell)$ of the correlation coefficient
maps of the ILC signals and synchrotron radiation.
Top to bottom: power spectra for the pixelization with the pixel
side of 160\arcmin, 300\arcmin and 540\arcmin. The dotted line, short and long
dashes mark the $\pm \sigma$, $\pm 2\sigma$ and $\pm$3$\sigma$
levels in the scatter of signal values for the correlation of a
random map and synchrotron radiation.}
\end{figure}

The spectral analysis shows that similarly to the case of dust,
there is a very notable harmonic $\ell=2$. In order to study its
orientation let us build its map (Fig.\,11).

\begin{figure} 
\centerline{\vbox{
\psfig{figure=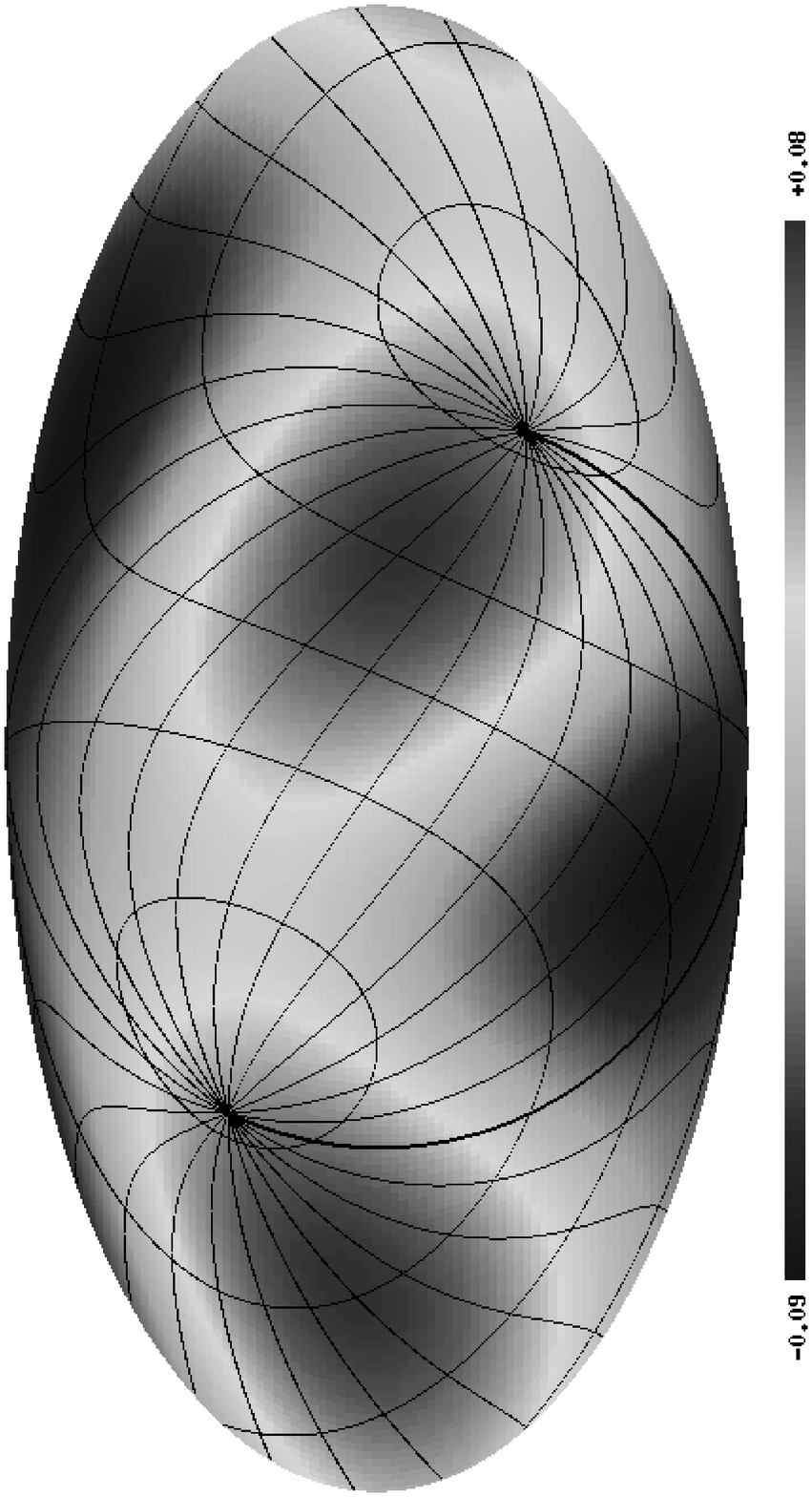,angle=-90,width=9cm}
\psfig{figure=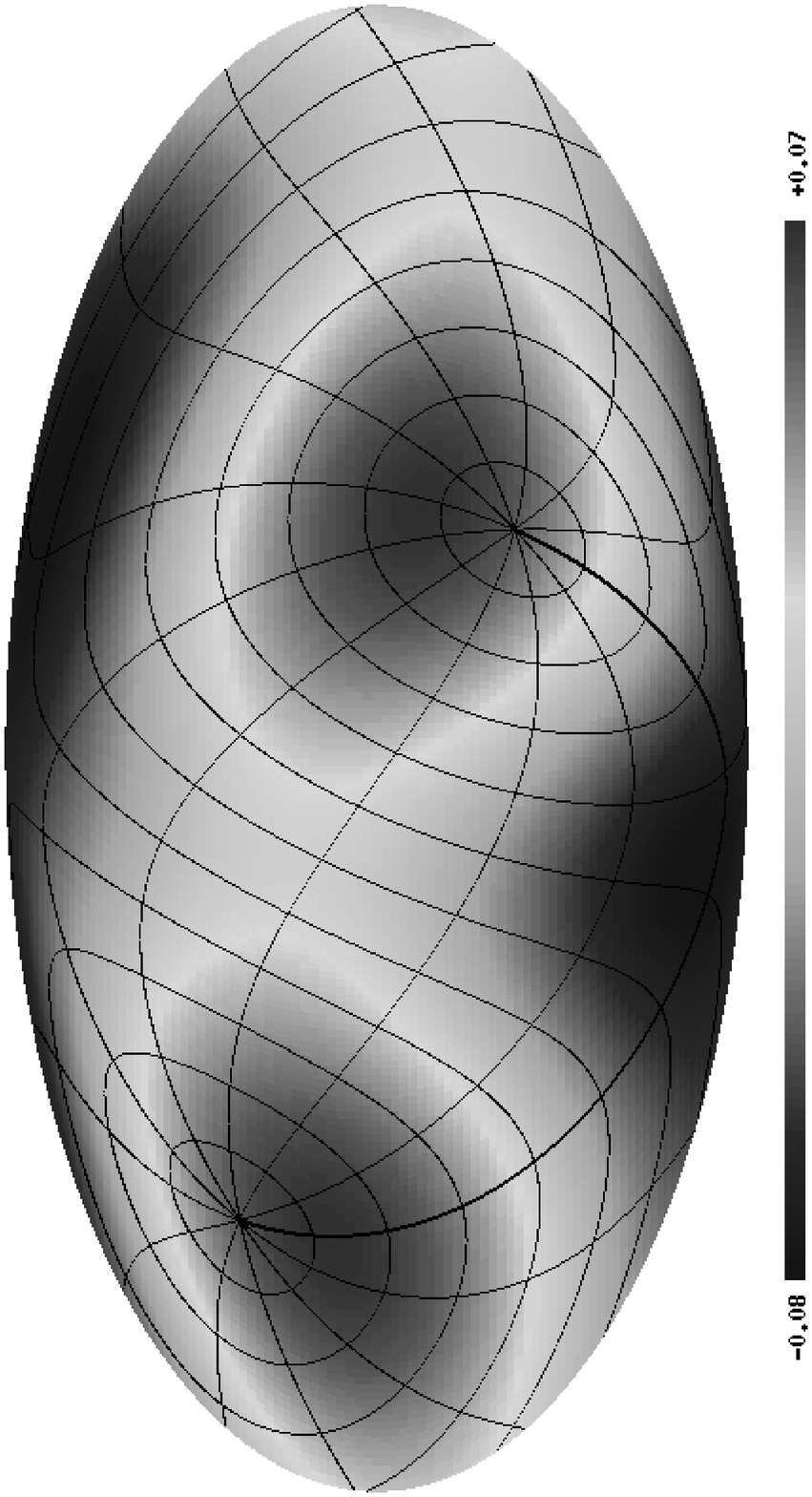,angle=-90,width=9cm}
\psfig{figure=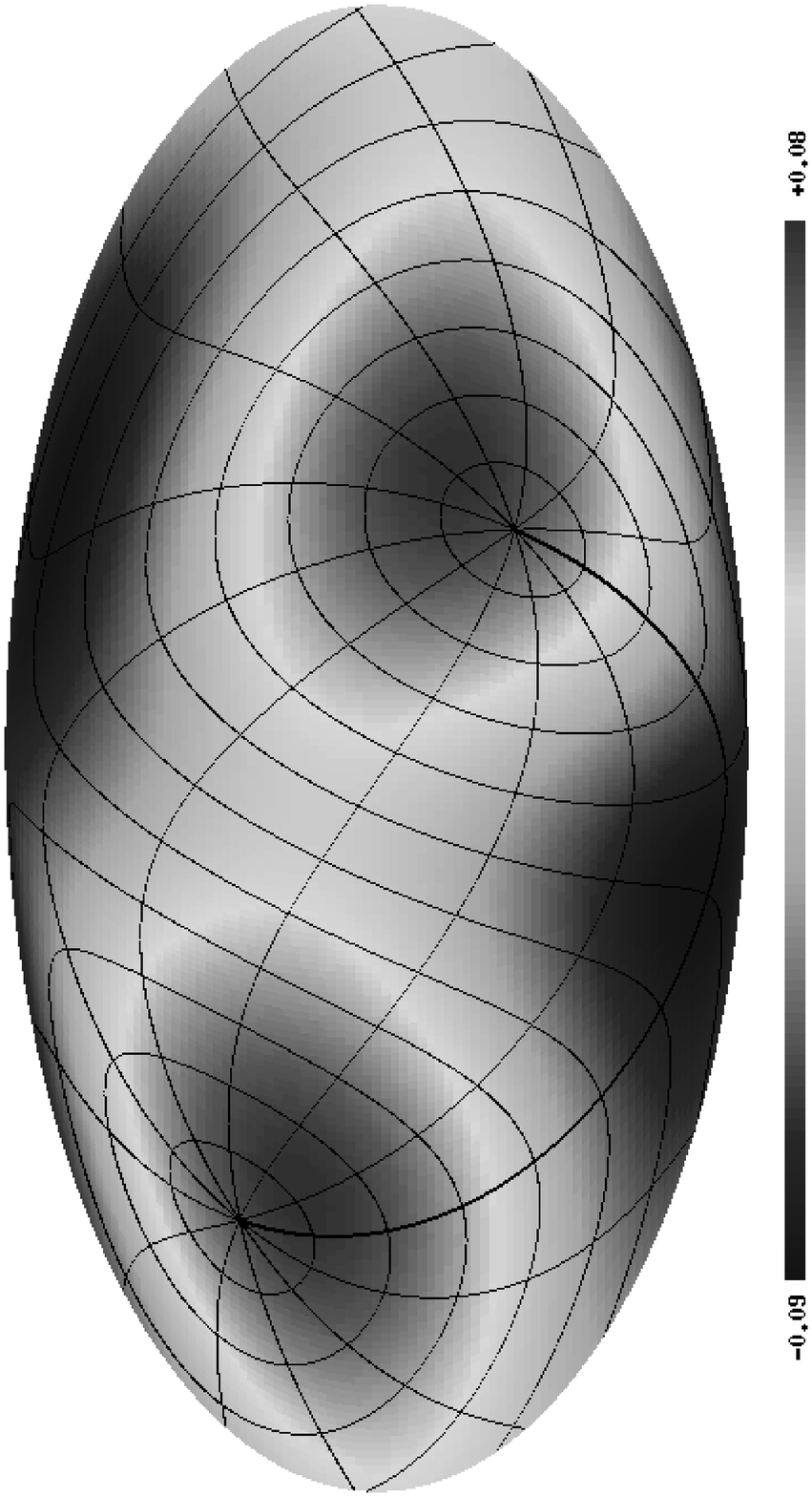,angle=-90,width=9cm}
}} \caption{The quadrupole, extracted from the ILC and synchrotron
radiation correlation maps.  Top to bottom: maps for the
pixelization of correlations within the areas with the sides of
160\arcmin, 300\arcmin and 540\arcmin. The first map has the ecliptic coordinate
grid superimposed, the second and the third---equatorial. The
coordinate grid poles fall on the hot spots (light gray), and the
cold spots (dark gray) lie in the equator.}
\end{figure}

Figure\,11 draws our attention to the location of a harmonic, the
orientation of which does not coincide with the Galactic plane.
Although hot spots are considerably shifted relative to the maxima
of the correlation maps with the dust (Fig.\,4), the positions of
cold spots (minima) indicate a notable sensitivity both to the
ecliptic plane (Fig.\,11, left, a quadrupole for the scale of
160\arcmin), and to the equatorial plane (Fig.\,11, right, a
quadrupole for the scale of 540\arcmin). Both minima in the two
maps lie at the corresponding zero latitude.

\section{DISCUSSION}

A stable presence of the quadrupole axes linking the hot and cold
spots, located in the same areas of the sky for different ILC and
dust component correlation map windows in conjunction with the
power spectrum, exceeding 1$\sigma$ manyfold, indicates a high
level of non-Gaussianity of low multipoles in the WMAP signal. A
distinctive feature of this paper from those of other authors as
well demonstrating the non-Gaussianity of the WMAP low multipoles
(see Introduction) is not only the method employed, but two more
components: (1) we show that when cleaning the WMAP data applying
the ILC technique, the dust component gives a strong
anticorrelation to the emitted CMB, manifested both in the
distribution of correlation coefficients, and in the angular power
spectrum. The synchrotron component is considerably manifested
within this approach  only in the power spectrum; (2) the
distribution of correlation coefficients allows speculating about
a sort of a special signal that has the ``knowledge'' of not only
the ecliptic coordinate system, but even more so about the
equatorial grid.

Note that two negative spots detected in the correlation map's
quadrupole are composed of negative values, indicating the inverse
behavior of the selected ILC map and background components in
these areas. The negative areas of the quadrupole strictly
correspond to the position of positive residual signal zones
detected in \cite{gruppuso2009:Verkh_ecl_en}. This fact confirms the prominence
of these WMAP map areas. An important argument from our point of
view is the position of these spots, which does not seem to be
entirely random. The coordinates of the minima of the quadrupole's
cold spots are close to the (0,0) and (180$^\circ$,0) coordinates
both in the ecliptic, and in the equatorial coordinate systems,
and the positions of hot spots (the poles either in the ecliptic,
or equatorial systems) are sensitive to the size of the selected
correlation scale. We have demonstrated this effect plotting a
grid on the quadrupole map (Fig.\,5 and 11).

Note that we can not yet confidently tell what has caused the
sensitivity to the coordinate system. We can enumerate several
potential effects that may play a role in enhancing the signal in
the ecliptic:
\begin{itemize}
\item the presence of the dust component in the ecliptic plane;
despite the weakness of contribution of the zodiacal dust
\cite{gruppuso2009:Verkh_ecl_en}, we \linebreak can not as yet
rule out the effect of silicone condensates
\cite{dikarev:Verkh_ecl_en}; \item the presence of the microwave
background, determined by the solar wind; the results of research
in this direction, presented in the form of signal distribution
maps on the full sphere, are not yet available; in addition, such
a background has to be variable owing to the solar activity; \item
the residual noise associated with the inhomogeneous sensitivity
of a WMAP pixel, depending on the distance from the ecliptic
plane; the produced mosaic correlation and modeling for the WMAP
noise map has shown very low correlation characteristics with the
ILC map (Fig.\,12);

\begin{figure*} 
\centerline{\hbox{
\psfig{figure=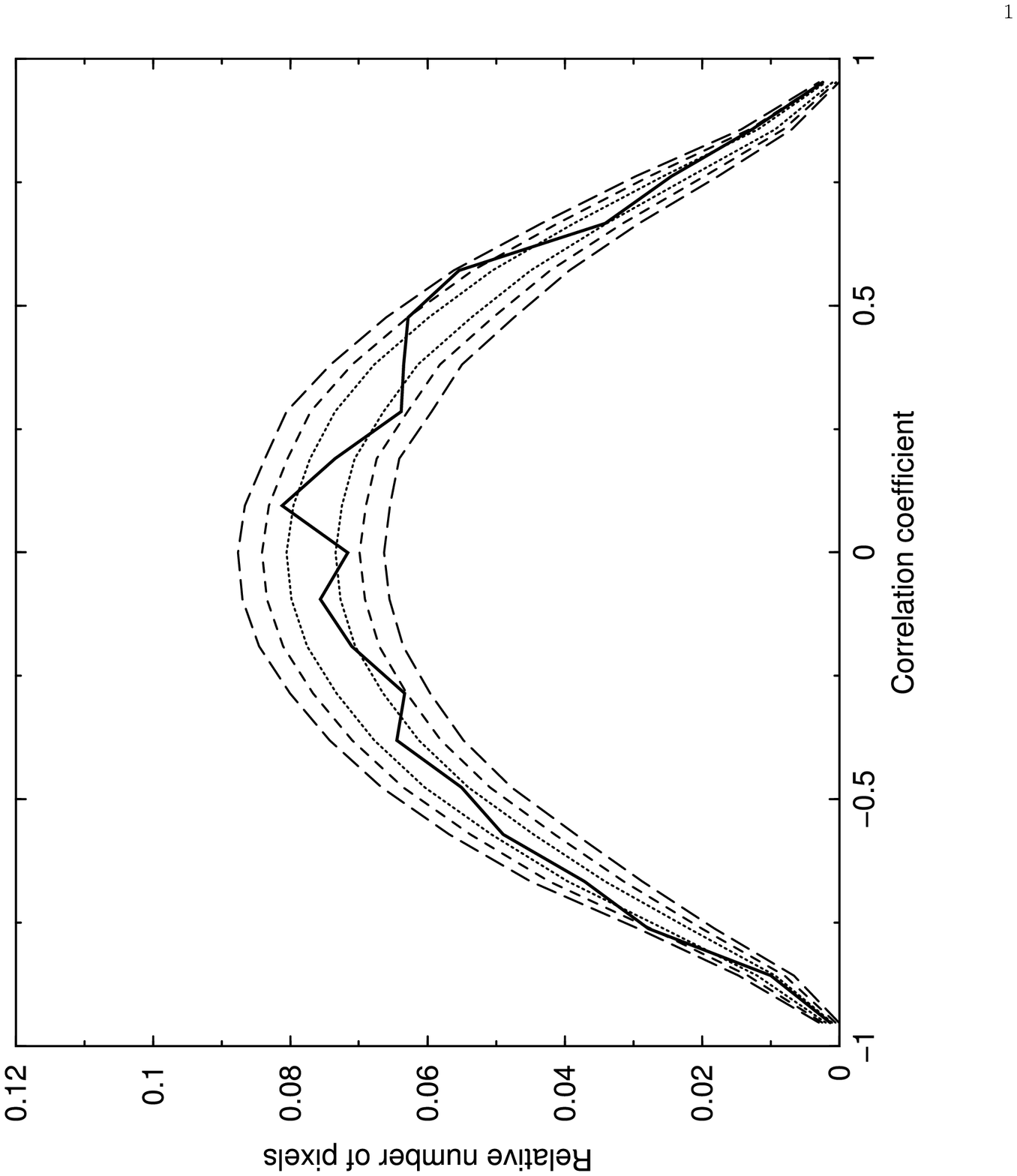,angle=-90,width=7cm}
\psfig{figure=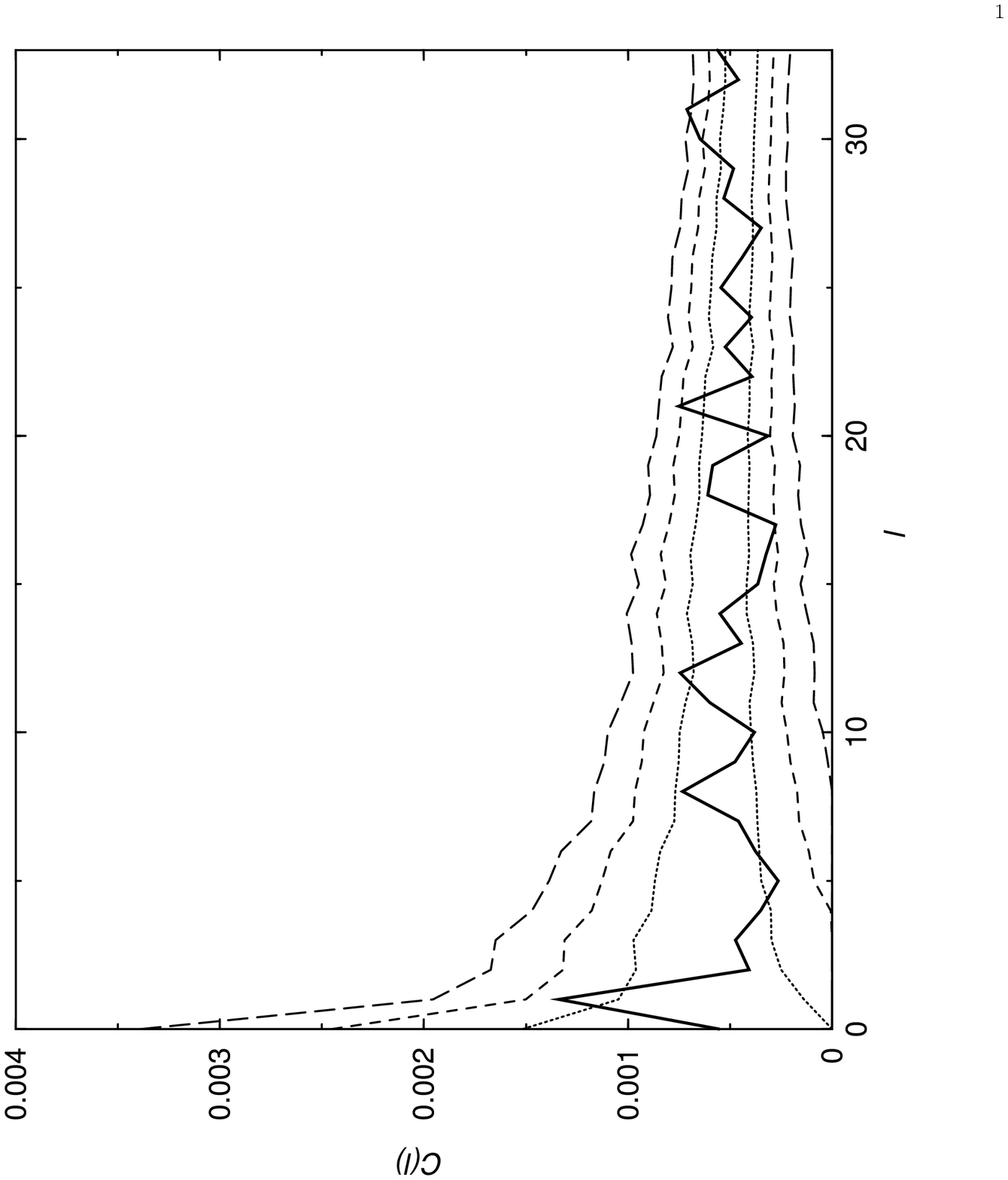,angle=-90,width=7cm}
}} \caption{Left: pixel histogram for the mosaic correlation map
of ILC and noise. Right: power spectrum of this map. The
correlations were computed inside the areas with the pixel side of
160\arcmin. The dotted line, short and long dashes mark the $\pm
\sigma$, $\pm 2\sigma$ and $\pm$3$\sigma$ levels in the pixel
distribution for correlating random and noise signals.}
\end{figure*}

\item The influence of huge dust belts of giant planets on the
effects of component separation of the microwave radiation, the
remnants of which are noticeable in the ecliptic plane. The giant
dust ring, newly discovered by NASA's Spitzer infrared space
telescope\footnote{{\tt http://science.nasa.gov/headlines/y2009/}
{\tt /07oct\_giantring.htm}} \cite{clavin:Verkh_ecl_en}, with the
size larger than one degree should give an unaccounted
contribution to the radiation in the ecliptic plane as a result of
multiple passes through the power beam of the WMAP antenna. The
level of such a contribution in the microwave range was not
evaluated.
\end{itemize}

A discovery of the equatorial coordinate system in the
correlations, apparent in the positions of the quadrupole hot and
cold spots, remains a surprise. The possibility of its physical
manifestation in the Lagrangian point L2 seems doubtful. There is
no evidence of the existence of the Earth's giant dust ring,
similar to those discovered around Saturn, as it would be
detectable both from Earth and from satellites. Neither do we find
any signatures of a giant tail of the Earth's magnetosphere
stretching up to L2, which could host areas of microwave
radiation. In this case, the effect of systematics, possibly
associated with the process of observations is probable, but not
clear. If we assume that non-Gaussian deviations of the low
multipole characteristics can be related to the problem of the
component separation, including the poorly-studied radiation of
the ecliptic plane, then the question of how these correlations
turn out to be sensitive to the equatorial system requires further
study.

The data anticipated from the Planck mission is of much interest
in this regard. These data, owing to higher resolution and
sensitivity, will allow to compute and investigate mosaic
correlation maps in the ecliptic and equatorial planes.

\noindent
{\small

\section{ACKNOWLEDGMENTS}

We are grateful to P.~D. Naselsky for useful discussions. We thank
the NASA for making available the NASA Legacy Archive, from where
we adopted the WMAP data. We are also grateful to the authors of
the HEALPix\footnote{\tt http://www.eso.org/science/healpix/}
\cite{healpix:Verkh_ecl_en} package, which we used to transform the WMAP maps
into the coefficients $a_{\ell m}$. This work made use of the
\mbox{GLESP\footnote{\tt http://www.glesp.nbi.dk}
\cite{glesp:Verkh_ecl_en,glesp1:Verkh_ecl_en}} package for the further analysis of the CMB
data on the sphere. This work was supported by the grant "Leading
Scientific Schools of Russia" and the Russian Foundation for Basic
Research (grant no.~09-02-00298). O.V.V. also acknowledges partial
support from the Foundation for the Support of Domestic Science
(the program ``Young Doctors of Science of the Russian Academy of
Sciences'') and the Dynasty Foundation.

}

\end{document}